\documentstyle[12pt,epsf]{article}

\advance \topmargin by -\headheight
\advance \topmargin by -\headsep

\evensidemargin \oddsidemargin
 \def\mod{{\hbox{\rm mod}}}

\def\ie{{\em i.e.}}
\def\ie{\hbox{\it i.e.}}

\def\CC{{\mathchoice
{\rm C\mkern-8mu\vrule height1.45ex depth-.05ex
width.05em\mkern9mu\kern-.05em}
{\rm C\mkern-8mu\vrule height1.45ex depth-.05ex
width.05em\mkern9mu\kern-.05em}
{\rm C\mkern-8mu\vrule height1ex depth-.07ex
width.035em\mkern9mu\kern-.035em}
{\rm C\mkern-8mu\vrule height.65ex depth-.1ex
width.025em\mkern8mu\kern-.025em}}}

\def\RR{{\rm I\kern-1.6pt {\rm R}}}

\def\ZZ{{\rm Z}\kern-3.8pt {\rm Z} \kern2pt}
\def\IB{\relax{\rm I\kern-.18em B}}
\def\ID{\relax{\rm I\kern-.18em D}}
\def\II{\relax{\rm I\kern-.18em I}}
\def\IP{\relax{\rm I\kern-.18em P}}

\def\np{Nucl. Phys.}
\def\pl{Phys. Lett.}
\def\prl{Phys. Rev. Lett.}
\def\pr{Phys. Rev.}

\def\jhep{J. High Energy Phys.}

\newcommand{\beq}{\begin{equation}}
\newcommand{\eeq}{\end{equation}}
\newcommand{\rc}{\nonumber\\}
\newcommand{\bear}{\begin{eqnarray}}
\newcommand{\eear}{\end{eqnarray}}

\def\to{\rightarrow}

\def\to{\rightarrow}

\newfont{\namefont}{cmr10}
\newfont{\addfont}{cmti7 scaled 1440}
\newfont{\boldmathfont}{cmbx10}
\newfont{\headfontb}{cmbx10 scaled 1728}
\renewcommand{\theequation}{{\rm\thesection.\arabic{equation}}}
\begin{document}
\begin{titlepage}

\begin{center} \Large \bf Supersymmetric defects in the Maldacena-N\'u\~nez 
background

\end{center}

\vskip 0.3truein
\begin{center}
Felipe Canoura${}^{\,*}$
\footnote{canoura@fpaxp1.usc.es},
Angel Paredes${}^{\,\dagger}$
\footnote{Angel.Paredes@cpht.polytechnique.fr}
and
Alfonso V. Ramallo${}^{\,*}$
\footnote{alfonso@fpaxp1.usc.es}

\vspace{0.3in}

${}^{\,*}$Departamento de F\'\i sica de Part\'\i culas, Universidade de
Santiago de Compostela \\and\\
Instituto Galego de F\'\i sica de Altas Enerx\'\i as (IGFAE)\\
E-15782 Santiago de Compostela, Spain
\vspace{0.3in}

${}^{\,\dagger}$ Centre de Physique Th\'eorique, \'Ecole Polytechnique, 91128
Palaiseau, France

\end{center}
\vskip 1
truein

\begin{center}
\bf ABSTRACT
\end{center}
We find supersymmmetric configurations of a D5-brane probe in the 
Maldacena-N\'u\~nez  background which are extended along one or two of the spatial directions 
of the gauge theory. These embeddings are worldvolume solitons which behave as codimension two
or one defects in the gauge theory and preserve two of the four supersymmetries of the
background.

\vskip4.6truecm
\leftline{US-FT-2/05}
\leftline{CPHT-RR 045.0705}
\leftline{hep-th/0507155 \hfill July 2005}
\smallskip
\end{titlepage}
\setcounter{footnote}{0}



\setcounter{equation}{0}
\section{Introduction}
\medskip
The AdS/CFT correspondence \cite{jm} is one of the most remarkable achievements of string theory
\cite{MAGOO}. In its original formulation  this correspondence states that ${\cal N}=4$ super
Yang-Mills theory in 3+1 dimensions is dual to type IIB supergravity on $AdS_5\times S^5$. By
adding additional structure to both sides of the correspondence one can get interesting
generalizations. In particular, on the field theory side one can add spatial defects which
reduce the amount of supersymmetry but nevertheless preserve conformal invariance \cite{dCFT},
giving rise to the so-called ``defect conformal field theories" (dCFT). 

A holographic dual of four-dimensional ${\cal N}=4$ super Yang-Mills theory with a
three-dimensional defect was proposed in ref. \cite{KR} by Karch and Randall, who conjectured
that such a dCFT can be realized in string theory by means of a D3-D5 intersection. In the
near-horizon limit the D3-branes give rise to an  $AdS_5\times S^5$ background, in which the
D5-branes wrap an $AdS_4\times S^2$ submanifold. It was argued in ref. \cite{KR} that the AdS/CFT
correspondence acts twice in this system and, apart from the holographic description of the four
dimensional field theory on the boundary of $AdS_5$, the fluctuations of the D5-brane should be
dual to the physics confined to the boundary  of $AdS_4$. In the probe
approximation the back-reaction of the D5-branes on the near-horizon geometry of the D3-branes is
neglected and the fluctuation modes of the $AdS_4$-brane are described by the Dirac-Born-Infeld
action of the D5-brane.

The defect conformal field theory associated with the D3-D5 intersection described above
corresponds to  ${\cal N}=4$, $d=4$ super Yang-Mills theory coupled to 
${\cal N}=4$, $d=3$ fundamental hypermultiplets localized at the defect. These hypermultiplets
arise as a consequence of the strings stretched between the D3- and D5-branes. In ref.
\cite{WFO}  the action of this model was constructed and a precise dictionary between operators
of the field theory and fluctuation modes of the probe was obtained (see also ref. \cite{EGK}).
The supersymmetry of the $AdS_4\times S^2\subset AdS_5\times S^5$ embedding of the D5-brane was
explicitly verified in ref. \cite{ST} by using kappa symmetry.

The defect field theories corresponding to other intersections have also been studied in the
literature. For example, from the D1-D3 intersection one gets a four-dimensional CFT with a
hypermultiplet localized on a one-dimensional defect \cite{1defect}. Moreover, the D3-D3
intersection gives rise to a two-dimensional defect in a four-dimensional CFT. In this case one
has, in the probe approximation, a D3-brane probe wrapping an  $AdS_3\times S^1$ submanifold of
the  $AdS_5\times S^5$ background. In ref. \cite{CEGK} the spectrum of fluctuations of the
D3-brane probe was obtained and the corresponding dual fields were identified. Let us finally
mention that the D3-D7 intersection leads to a  configuration in which the D7-brane fills
completely the four-dimensional spacetime
(a codimension zero ``defect") which has been used to
add flavor to  ${\cal N}=4$ super Yang-Mills theory \cite{D3D7,BDFLM,KK,KMMW}.

The extension of the above ideas to more realistic theories is of great interest. It is
important to recall in this respect that four 
dimensional gauge theories with ${\cal N}=1$ supersymmetry
share some qualitative features with the physics of the real world such as,
for instance, confinement. One can  think of
modifying the theory by introducing defects (regions of space-time where some fields are
localized), which break the $SO(1,3)$ Lorentz invariance. 

Actually, 
the superalgebras of such field theories
admit central charges associated with objects extended in two
or one space directions (codimension one and two, respectively).
For instance, in $SU(N)$ ${\cal N}=1$ SYM there are 1/2-BPS domain
walls which interpolate between the inequivalent $N$ vacua which
come from the spontaneous breaking of the $\ZZ_{2N}$ symmetry (the
non-anomalous subgroup of the $U(1)_R$) to $\ZZ_2$ by the gaugino condensate.
There can be also BPS codimension two objects, namely strings (flux 
tubes) which have been studied in the context of different ${\cal N}=1$ 
theories, see \cite{Gorsky:1999hk} and references therein. For recent progress along 
these lines, see also \cite{Shifman:2005st} and references therein. The physics of such 
objects turns out to be quite rich, including for instance the phenomenon 
of enhanced (supersized) supersymmetry, also present for domain 
walls \cite{Shifman:2005st}.  Holographically, these objects 
can be introduced by adding
D-branes to the setup. As argued above, one can also think of adding
supersymmetric defects also of codimension one or two. Since this
modifies the lagrangian of the field theory, we expect, on general 
grounds, that the string theory setup should be modified at infinity.
Therefore, the defects should be dual to D-branes extending infinitely
in the holographic direction.

The purpose of this work is to make a rather systematic search for
possible supersymmetric embeddings for D-brane probes in a concrete
model, the Maldacena-Nu\~nez  dual \cite{MN,CV} to $SU(N)$ ${\cal N}=1$ super Yang-Mills (for a
review see \cite{MerReview}). As shown in ref. 
\cite{flavoring}, the Maldacena-Nu\~nez  background  has a rich structure of
submanifolds along which one can wrap a D5-brane probe without breaking supersymmetry
completely. Actually, a series of such embeddings in which the D5-brane probe fills all the
gauge theory directions was found in ref. \cite{flavoring}. In those configurations the probe
preserves the same supersymmetry as the background and some of them can be considered as flavor
branes \cite{WangHu}, suitable to study the meson spectrum of four-dimensional ${\cal N}=1$ super
Yang-Mills theory.

In this paper we continue the analysis of \cite{flavoring} by studying the configurations of
D5-brane probes which are a codimension one or two defect in the gauge theory directions. As in 
ref. \cite{flavoring} the main tool used will be kappa symmetry \cite{bbs}, which is based on
the fact that there exists certain matrix $\Gamma_{\kappa}$ such that, if $\epsilon$ is a
Killing spinor of the background, only those embeddings satisfying
$\Gamma_{\kappa}\,\epsilon=\epsilon$ preserve some supersymmetry \cite{swedes}. The matrix 
$\Gamma_{\kappa}$ depends on the metric induced on the worldvolume of the probe and, actually, by
imposing the equation $\Gamma_{\kappa}\,\epsilon=\epsilon$ one can systematically determine the
supersymmetric embeddings of the  probe and it is possible to identify the fraction of
supersymmetry  preserved by the configuration.

Obviously, to apply the technique sketched above one has to know first the Killing spinors
of the background. For the Maldacena-Nu\~nez solution these spinors were determined in 
\cite{flavoring}. It will become clear from these  spinors that only the D5-brane probes can 
have supersymmetric embeddings of the type we are interested in  here. The configurations we will
find  preserve two of the four supersymmetries of the background. Of course, it would be 
interesting to find  the backreaction in the
geometry of the branes described above, but this is  a technically challenging
problem and is beyond the scope of the present work.

This paper is organized as follows. In section 2 we review the main properties of the 
Maldacena-Nu\~nez solution which we will need in the paper. In this section we will also recall
the kappa symmetry condition for supersymmetric probes and we will discuss the general strategy
to solve it. In section 3 we will find D5-brane embeddings which are  wall defects from the gauge
theory point of view. In section 4 we will obtain supersymmetric configurations in which the
D5-brane probe is extended only along one of the spatial gauge theory directions and behaves as
a codimension two defect of the field theory. Section 5 is devoted to summarize our results and
to some concluding remarks.  In appendix A we will demonstrate that the
solutions found in sections 2 and 3 saturate  certain energy bound. In appendix B we discuss
other  wall defect solutions different from those found in the main text.

\setcounter{equation}{0}
\section{The Maldacena-Nu\~nez solution}
\medskip
The Maldacena-Nu\~nez (MN) background is a solution of the equations of motion of
type IIB supergravity which preserves four supersymmetries. It can be obtained \cite{MN, CV}
as a solution of seven-dimensional gauged supergravity,  which is a consistent truncation of
ten-dimensional supergravity. The seven-dimensional solution is subsequently uplifted to ten
dimensions. The ten-dimensional metric  in the string frame is:
\beq
ds^2_{10}\,=\,e^{\phi}\,\,\Big[\,
dx^2_{1,3}\,+\,e^{2h}\,\big(\,d\theta_1^2+\sin^2\theta_1 d\phi_1^2\,\big)\,+\,
dr^2\,+\,{1\over 4}\,(w^i-A^i)^2\,\Big]\,\,,
\label{metric}
\eeq
where $\phi$ is the dilaton, $h$ is a function which depends on the radial coordinate $r$, the
one-forms $A^i$ $(i=1,2,3)$ are
\beq
A^1\,=\,-a(r) d\theta_1\,,
\,\,\,\,\,\,\,\,\,
A^2\,=\,a(r) \sin\theta_1 d\phi_1\,,
\,\,\,\,\,\,\,\,\,
A^3\,=\,- \cos\theta_1 d\phi_1\,,
\label{oneform}
\eeq
and the $w^i$'s are  $su(2)$ left-invariant one-forms,
satisfying  $dw^i=-{1\over 2}\,\epsilon_{ijk}\,w^j\wedge w^k$. The $A^i$'s are the components of
the non-abelian gauge vector field of the seven-dimensional gauged supergravity. Moreover, 
the $w^i$'s parametrize the
compactification three-sphere and can be represented in
terms of three angles $\phi_2$, $\theta_2$ and $\psi$:
\bear
w^1&=& \cos\psi d\theta_2\,+\,\sin\psi\sin\theta_2
d\phi_2\,\,,\rc\rc
w^2&=&-\sin\psi d\theta_2\,+\,\cos\psi\sin\theta_2 
d\phi_2\,\,,\rc\rc
w^3&=&d\psi\,+\,\cos\theta_2 d\phi_2\,\,.
\eear
The angles $\theta_i$, $\phi_i$ and $\psi$ take values in the intervals $\theta_i\in [0,\pi]$, 
$\phi_i\in [0,2\pi)$ and $\psi\in [0,4\pi)$. The functions $a(r)$, $h(r)$ and the dilaton $\phi$
are:
\bear
a(r)&=&{2r\over \sinh 2r}\,\,,\rc\rc
e^{2h}&=&r\coth 2r\,-\,{r^2\over \sinh^2 2r}\,-\,
{1\over 4}\,\,,\rc
e^{-2\phi}&=&e^{-2\phi_0}{2e^h\over \sinh 2r}\,\,.
\label{MNsol}
\eear

The solution of the type IIB supergravity also includes a Ramond-Ramond three-form $F_{(3)}$
given by
\beq
F_{(3)}\,=\,-{1\over 4}\,\big(\,w^1-A^1\,\big)\wedge 
\big(\,w^2-A^2\,\big)\wedge \big(\,w^3-A^3\,\big)\,+\,{1\over 4}\,\,
\sum_a\,F^a\wedge \big(\,w^a-A^a\,\big)\,\,,
\label{RRthreeform}
\eeq
where $F^a$ is the field strength of the su(2) gauge field $A^a$, defined as $
F^a\,=\,dA^a\,+\,{1\over 2}\epsilon_{abc}\,A^b\wedge A^c$.

In order to write the Killing spinors of the background in a simple form, let us consider the
frame:
\bear
e^{x^i}&=&e^{{\phi\over 2}}\,d x^i\,\,,
\,\,\,\,\,\,\,(i=0,1,2,3)\,\,,\rc\rc
e^{1}&=&e^{{\phi\over 2}+h}\,d\theta_1\,\,,
\,\,\,\,\,\,\,\,\,\,\,\,\,\,
e^{2}=e^{{\phi\over 2}+h}\,\sin\theta_1 d\phi_1\,\,,\rc\rc
e^{r}&=&e^{{\phi\over 2}}\,dr\,\,,
\,\,\,\,\,\,\,\,\,\,\,\,\,\,
e^{\hat i}={e^{{\phi\over 2}}\over 2}\,\,
(\,w^i\,-\,A^i\,)\,\,,\,\,\,\,\,\,\,(i=1,2,3)\,\,.
\label{frame}
\eear
Let $\Gamma_{x^i}$ ($i=0,1,2,3$), $\Gamma_{j}$ ($j=1,2$), $\Gamma_{r}$ and
$\hat\Gamma_{k}$ ($k=1,2,3$) be constant Dirac matrices associated to the frame
(\ref{frame}). Then, the Killing spinors of the MN solution satisfy \cite{flavoring}:
\bear
&&\Gamma_{x^0\cdots x^3}\,\Gamma_{12}\,\epsilon\,=\,
\Gamma_{r}\hat \Gamma_{123}\,\epsilon\,=\,e^{-\alpha\Gamma_1\hat\Gamma_1}
\epsilon\,=\,\big[\,\cos\alpha\,-\,\sin\alpha\Gamma_1\hat\Gamma_1\,\big]\,
\epsilon\,\,,\rc\rc
&&\Gamma_{12}\,\epsilon\,=\,\hat\Gamma_{12}\,\epsilon\,\,,\rc\rc
&&\epsilon\,=\,i\epsilon^*\,\,,
\label{fullprojection}
\eear
where the angle $\alpha$ is given by
\beq
\sin\alpha\,=\,-{ae^h\over r}\,\,,
\,\,\,\,\,\,\,\,\,\,\,\,
\cos\alpha\,=\,{e^{2h}\,-\,{1\over 4}\,(\,a^2-1\,)\over r}\,\,.
\label{alpha}
\eeq
A simple expression for $\cos\alpha$ as a function of $r$ can be written, namely
\beq
\cos\alpha\,=\,{\rm \coth} 2r\,-\,{2r\over \sinh^22r}\,\,.
\label{alphaexplicit}
\eeq
In the first equation in (\ref{fullprojection}) we have used the fact that $\epsilon$ is a
spinor of definite chirality. Moreover, from the above equations we can obtain the explicit form
of the Killing spinor $\epsilon$. It can be written as:
\beq
\epsilon\,=\,f(r)\,e^{{\alpha\over 2}\,\Gamma_1\hat\Gamma_1}\,\,\,\eta\,\,,
\label{epsiloneta}
\eeq
where $f(r)$ is a commuting function of the radial coordinate, whose explicit expression is
irrelevant in what follows, and $\eta$ is a constant spinor which satisfies:
\beq
\Gamma_{x^0\cdots x^3}\,\Gamma_{12}\,\eta\,=\,\eta\,\,,
\,\,\,\,\,\,\,\,\,\,\,\,
\Gamma_{12}\,\eta\,=\,\hat\Gamma_{12}\,\eta\,\,,
\,\,\,\,\,\,\,\,\,\,\,\,
\eta\,=\,i\eta^*\,\,.
\label{constantfullpro}
\eeq

Apart from the full regular MN solution described above we shall also consider the simpler
background in which the function $a(r)$ vanishes and, thus, the one-form $A$ has only one
non-vanishing component, namely $A^3$. This solution is singular in the IR and coincides with
the regular MN background in the UV region $r\to\infty$. Indeed, by taking $r\to\infty$ in the
expression of $a(r)$ in eq. (\ref{MNsol}) one gets $a(r)\to 0$. Moreover, by neglecting
exponentially suppressed terms one gets:
\beq
e^{2h}\,=\,r\,-\,{1\over 4}\,\,,
\,\,\,\,\,\,\,\,\,\,\,\,\,\,\,\,\,\,(a=0)\,\,,
\label{abelianh}
\eeq
while $\phi(r)$ can be obtained by using the expression of $h$ given in eq. (\ref{abelianh}) on
the last equation in (\ref{MNsol}). The RR three-form $F_{(3)}$ is still given by eq.
(\ref{RRthreeform}), but now $A^1=A^2=0$ and $A^3$ is the same as in eq. (\ref{oneform}). 
We will refer to this solution as the abelian MN background. The metric of this abelian MN
background is singular at $r={1\over 4}$ (by redefining the radial coordinate this singularity
could be moved to $r=0$). Moreover, the Killing spinors in this abelian case can be obtained
from those of the regular background by simply putting $\alpha=0$, which is indeed the value
obtained by taking the $r\to\infty$ limit on the right-hand side of eq. (\ref{alpha}).

Since $dF_{(3)}=0$, one can find a two-form potential $C_{(2)}$ such that 
$F_{(3)}=dC_{(2)}$. The expression of $C_{(2)}$, which will not be needed here, can be found
in ref. \cite{flavoring}. Moreover, the equation of motion satisfied by $F_{(3)}$ is
$d{}^*F_{(3)}=0$, where $*$ denotes Hodge duality. Therefore one can write, at least locally, 
${}^*F_{(3)}\,=\,d C_{(6)}$, with  $C_{(6)}$ being a six-form potential. The expression of 
$C_{(6)}$ can be taken from the results of ref. \cite{flavoring}, namely:
\beq
C_{(6)}\,=\,dx^0\wedge dx^1\wedge dx^2\wedge dx^3\wedge
{\cal C}\,\,,
\label{C6}
\eeq
where ${\cal C}$ is the following two-form:
\bear
{\cal C}&=&-{e^{2\phi}\over 8}\,\,\Big[\,
\Big(\,(\,a^2-1\,)a^2\,e^{-2h}\,-\,16\,e^{2h}\,\Big)\,\cos\theta_1
d\phi_1\wedge dr
\,-\,(\,a^2-1\,)\,e^{-2h}\,w^3\wedge dr\,+\rc\rc
&&+\,a'\,\Big(\,\sin\theta_1 d\phi_1\wedge w^1\,+\,d\theta_1\wedge
w^2\,\Big)\,\Big]\,\,.
\label{calC}
\eear 

It is also interesting to recall the isometries of the abelian and non-abelian metrics. In the
abelian solution $a=0$ the angle $\psi$ does not appear in the expression of the metric 
(\ref{metric}) (only $d\psi$ does). Therefore, $\psi$ can be shifted by an arbitrary constant
$\lambda$ as $\psi\to\psi+\lambda$. Actually, this $U(1)$ isometry of the abelian metric is
broken quantum-mechanically to a $\ZZ_{2N}$ subgroup as a consequence of the flux quantization
condition of the RR two-form potential \cite{MN,anomaly,BDFLM}.  In the gauge theory side this
isometry can be identified with the $U(1)$ R-symmetry, which is broken in the UV to the same 
$\ZZ_{2N}$ subgroup by a field theory anomaly. On the contrary, the non-abelian metric does
depend on
$\psi$ through $\sin\psi$ and $\cos\psi$ and, therefore, only the discrete $\ZZ_2$ isometry
$\psi\to\psi+2\pi$ remains when $a\not=0$. This fact has been interpreted \cite{MN, Apreda} as
the string theory dual of the spontaneous breaking of the R-symmetry induced by the gluino
condensate in the IR.

\subsection{Supersymmetric Probes}
We will now consider a Dp-brane probe embedded in the MN geometry (\ref{metric}). If we denote
by  $\xi^{\mu}$ ($\mu=0,\cdots,p$) a set of worldvolume coordinates and if $X^M$ are
ten-dimensional coordinates, the embedding of the Dp-brane is determined by a set of functions
$X^M(\xi^{\mu})$. The induced metric on the worldvolume is defined as:
\beq
g_{\mu\nu}\,=\,\partial_{\mu} X^{M}\,\partial_{\nu} X^{N}\,G_{MN}\,\,,
\label{inducedmetric}
\eeq
where $G_{MN}$ is the ten-dimensional metric (\ref{metric}). Moreover, let $e^{\underline{M}}$
denote the one-forms of the frame basis (\ref{frame}). The $e^{\underline{M}}$ one-forms
can be written in terms of the differentials of the coordinates by means of the vielbein
coefficients
$E_{N}^{\underline{M}}$, namely:
\beq
e^{\underline{M}}\,=\,E_{N}^{\underline{M}}\,dX^N\,\,.
\eeq
Then, the induced Dirac matrices on
the worldvolume are defined as
\beq
\gamma_{\mu}\,=\,\partial_{\mu}\,X^{M}\,E_{M}^{\underline{N}}\,\,
\Gamma_{\underline{N}}\,\,,
\label{wvgamma}
\eeq
where $\Gamma_{\underline{N}}$ are constant ten-dimensional Dirac matrices. The supersymmetric
BPS configurations of the brane probe are obtained by imposing the condition:
\beq
\Gamma_{\kappa}\,\epsilon\,=\,\epsilon\,\,,
\label{kappacondition}
\eeq
where $\Gamma_{\kappa}$ is a matrix which depends on the embedding of the probe (see below) and
$\epsilon$ is a Killing spinor of the background. In order to write the expression of
$\Gamma_{\kappa}$ it is convenient to decompose the complex spinor $\epsilon$ used up to now in
its real and imaginary parts as $\epsilon\,=\,\epsilon_1+i\epsilon_2$. We shall arrange the two
Majorana-Weyl spinors $\epsilon_1$ and $\epsilon_2$ as a two-dimensional vector 
$\pmatrix{\epsilon_1\cr\epsilon_2}\,\,$. It is
straightforward to find the following rules to pass from complex to real spinors:
\beq
\epsilon^*\,\leftrightarrow\,\tau_3\,\epsilon\,\,,
\,\,\,\,\,\,\,\,\,\,\,\,\,\,\,\,\,\,\,
i\epsilon^*\,\leftrightarrow\,\tau_1\,\epsilon\,\,,
\,\,\,\,\,\,\,\,\,\,\,\,\,\,\,\,\,\,\,
i\epsilon\,\leftrightarrow\,-i\tau_2\,\epsilon\,\,,
\label{rule}
\eeq
where the $\tau_i$ $(i=1,2,3)$ are Pauli matrices that act on the two-dimensional vector 
$\pmatrix{\epsilon_1\cr\epsilon_2}$.

We will assume that there  are not worldvolume gauge fields on the D-brane, which is consistent
with the equations of motion of the probe if there are not source terms which could induce them.
These source terms must be linear in the gauge field and can only be originated in the
Wess-Zumino part of the probe action. For the cases considered below we will verify that the RR
potentials of the MN background do not act as source of the worldvolume gauge fields and,
therefore, the latter can be consistently put to zero. If this is the case, the kappa symmetry
matrix of a Dp-brane in the type IIB theory,  acting on the real two-component
spinors, is given by \cite{swedes}:
\beq
\Gamma_{\kappa}\,=\,{1\over (p+1)!\sqrt{-g}}\,\epsilon^{\mu_1\cdots\mu_{p+1}}\,
(\tau_3)^{{p-3\over 2}}\,i\tau_2\,\otimes\,
\gamma_{\mu_1\cdots\mu_{p+1}}\,\,,
\label{gammakappa}
\eeq
where $g$ is the determinant of the induced metric $g_{\mu\nu}$ and
$\gamma_{\mu_1\cdots\mu_{p+1}}$ denotes the  antisymmetrized product of the induced gamma
matrices.  

The kappa symmetry equation $\Gamma_{\kappa}\,\epsilon\,=\,\epsilon$ imposes a condition on the
Killing spinors which should be compatible with the ones required by the supersymmetry of the
background. These latter conditions are precisely the ones written in eq.
(\ref{fullprojection}). In particular (see eq. (\ref{fullprojection})) the spinor $\epsilon$ must
be such that $\epsilon=i\epsilon^*$, which in the real notation is equivalent to 
$\tau_1\epsilon=\epsilon$. Notice that the Pauli matrix appearing in the
expression of $\Gamma_{\kappa}$ in  (\ref{gammakappa}) is $\tau_1$ or $\tau_2$, depending on the
dimensionality of the probe. Clearly, the conditions $\Gamma_{\kappa}\,\epsilon\,=\,\epsilon$ and
$\tau_1\epsilon=\epsilon$ can only  be compatible if $\Gamma_{\kappa}$ contains the Pauli matrix
$\tau_1$. By inspecting eq. (\ref{gammakappa}) one readily realizes that this happens for
$p=1,5$.  Moreover, we want our probes to be extended both along the spatial Minkowski and
internal directions, which is not possible for Lorentzian D1-branes and leaves us with the
D5-branes as the only case to be studied.  Notice that for the MN background the only couplings
of the Wess-Zumino term of the action linear in the worldvolume gauge field $F$ are of the form
$C^{(2)}\wedge F$ and $C^{(6)}\wedge F$, where $C^{(2)}$ and $C^{(6)}$ are the RR potentials. By
simple counting of the degree of these forms one immediately concludes that these terms are not
present in the action of a D5-brane and, thus, the gauge fields can be consistently taken to be
zero, as claimed above.

Coming back to the complex notation for the spinors, and taking into account the fact that the
Killing spinors of the MN background satisfy the condition $\epsilon=i\epsilon^*$, one can write
the matrix $\Gamma_{\kappa}$ for a D5-brane probe as:
\beq
\Gamma_{\kappa}\,=\,{1\over 6!}\,\,{1\over \sqrt{-g}}\,\,
\epsilon^{\mu_1\cdots \mu_6}\,\,\gamma_{\mu_1\cdots \mu_6}\,\,.
\label{GammaD5}
\eeq

Notice that, for a general embedding, the kappa symmetry condition
$\Gamma_{\kappa}\,\epsilon=\epsilon$ imposes a new projection to the Killing spinor $\epsilon$.
This new projection is not, in general, consistent with the conditions (\ref{fullprojection}),
since it involves matrices which do not commute with those appearing in (\ref{fullprojection}).
The only way of making the equation $\Gamma_{\kappa}\,\epsilon=\epsilon$ and
(\ref{fullprojection}) consistent with each other is by requiring the vanishing of the
coefficients of those non-commuting matrices. On the contrary, the terms in $\Gamma_{\kappa}$
which commute with the projections (\ref{fullprojection}) should act on the Killing spinors as
the unit matrix. These conditions will give rise to a set of first-order
BPS differential equations. By solving these BPS equations we will determine the
embeddings of the D5-brane we are interested in, namely those which preserve some
fraction of the background supersymmetry.

\setcounter{equation}{0}
\section{Wall defects}
\medskip
In this section we are going to find supersymmetric configurations of a D5-brane probe which,
from the point of view of the four-dimensional gauge theory, are codimension one objects.
Accordingly, we extend the D5-brane along three of the Minkowski coordinates $x^{\mu}$ (say 
$x^0$, $x^1$, $x^2$) and along a three dimensional submanifold of the internal part of the
metric. To describe these configurations it is convenient to choose 
the following set of worldvolume coordinates:
\beq
\xi^{m}\,=\,(x^0,x^1,x^2,r,\theta_1,\phi_1)\,\,.
\label{DWvwcoordinates}
\eeq
Moreover, we will adopt the following ansatz for the dependence of the remaining
ten-dimensional coordinates on the $\xi^{\mu}$'s:
\bear
&&x^3=x^3(r),\,\,\rc
&&\theta_2=\theta_2(\theta_1,\phi_1)\,\,,
\,\,\,\,\,\,\,\,\,\,\,\,\,
\phi_2=\phi_2(\theta_1,\phi_1)\,\,,\rc
&&\psi=\psi_0={\rm constant}\,\,,
\label{DWansatz}
\eear
In the appendix B we will explore other possibilities and, in particular, we will study
configurations for which $\psi$ is not constant. For the set of worldvolume coordinates
(\ref{DWvwcoordinates}) the kappa symmetry matrix acts on the Killing spinors $\epsilon$ as:
\beq
\Gamma_{\kappa}\,\epsilon\,=\,{1\over \sqrt{-g}}\,
\gamma_{x^0 x^1x^2 r\theta_1\phi_1}\,\epsilon\,\,.
\label{DWGammak}
\eeq
The induced gamma matrices appearing on the right-hand side of eq. (\ref{DWGammak}) can be
straightforwardly computed from the general expression (\ref{wvgamma}). One gets:
\bear
&&e^{-{\phi\over 2}}\,\gamma_{x^{\mu}}=\,\Gamma_{x^{\mu}}\,\,,
\,\,\,\,\,\,\,\,\,\,\,\,\,\,\,\,\,\,
(\mu=0,1,2),\,\,\rc\rc
&&e^{-{\phi\over 2}}\,\gamma_r\,=\,\Gamma_r\,+\,\partial_rx^3\Gamma_{x^3}\,\,,\rc\rc
&&e^{-{\phi\over 2}}\,\gamma_{\theta_1}\,=\,e^{h}\Gamma_1\,+\,
\big(\,V_{1\theta}+{a\over 2}\,\big)\,\hat\Gamma_1\,+\,V_{2\theta}\,\hat\Gamma_2\,+\,
V_{3\theta}\,\hat\Gamma_3\,\,,\rc
&&{e^{-{\phi\over 2}}\over \sin\theta_1}\,\gamma_{\phi_1}\,=\,e^{h}\Gamma_2\,+\,
V_{1\phi}\,\hat\Gamma_1\,+\,
\big(\,V_{2\phi}-{a\over 2}\,\big)\,\hat\Gamma_2\,+\
V_{3\phi}\,\hat\Gamma_3\,\,,
\label{DWDiracma}
\eear
where the $V$'s are the quantities:
\bear
&&V_{1\theta}\equiv {1\over 2}\,\Big[\cos\psi_0\,\partial_{\theta_1}\theta_2\,+\,
\sin\psi_0\,\sin\theta_2\,\partial_{\theta_1}\phi_2\,\Big]\,\,,\rc\rc
&&V_{2\theta}\equiv {1\over 2}\,\Big[-\sin\psi_0\,\partial_{\theta_1}\theta_2\,+\,
\cos\psi_0\,\sin\theta_2\,\partial_{\theta_1}\phi_2\,\Big]\,\,,\rc\rc
&&V_{3\theta}\equiv {1\over 2}\,\cos\theta_2\,\partial_{\theta_1}\phi_2\,\,,\rc\rc
&&\sin\theta_1V_{1\phi}\equiv {1\over 2}\,\Big[\cos\psi_0\,
\partial_{\phi_1}\theta_2\,+\,
\sin\psi_0\,\sin\theta_2\,\partial_{\phi_1}\phi_2\,\Big]\,\,,\rc\rc
&&\sin\theta_1\,V_{2\phi}\equiv {1\over 2}\,\Big[-\sin\psi_0\,
\partial_{\phi_1}\theta_2\,+\,
\cos\psi_0\,\sin\theta_2\,\partial_{\phi_1}\phi_2\,\Big]\,\,,\rc\rc
&&\sin\theta_1\,V_{3\phi}\equiv {1\over 2}\,\Big[\cos\theta_1\,+\,
\cos\theta_2\,\partial_{\phi_1}\phi_2\,\Big]\,\,.
\label{Vs}
\eear
Notice that the $V$'s depend on the angular part of the embedding (\ref{DWansatz}), \ie\ on the
functional dependence of $\theta_2$, $\phi_2$ on $(\theta_1, \phi_1)$. Using the expressions of
the $\gamma$'s given in eq. (\ref{DWDiracma}), one can write the action of $\Gamma_{\kappa}$ on 
$\epsilon$ as:
\beq
\Gamma_{\kappa}\,\epsilon\,=\,{e^{2\phi}\over \sqrt{-g}}\,
\Gamma_{x^0x^1x^2}\,\big[\,\Gamma_r\,+\,\partial_rx^3\Gamma_{x^3}\,\big]\,
\gamma_{\theta_1\phi_1}\,\epsilon\,\,.
\eeq
Moreover, by using the projection $\Gamma_{12}\,\epsilon\,=\,\hat\Gamma_{12}\,\epsilon$ (see eq.
(\ref{fullprojection})), $\gamma_{\theta_1\phi_1}\,\epsilon$ can be written as:
\bear
&&{e^{-\phi}\over \sin\theta_1}\,\gamma_{\theta_1\phi_1}\,\epsilon=\,
\big[\,
c_{12}\,\Gamma_{12}\,+\,c_{1\hat 2}\,\Gamma_{1}\hat\Gamma_{2}\,+\,
c_{1\hat 1}\,\Gamma_{1}\hat\Gamma_{1}\,+\,
\rc\rc
&&\,\,\,\,\,\,\,\,\,\,\,\,\,\,\,\,\,\,\,\,\,\,\,\,\,\,\,\,\,\,\,\,\,\,\,\,\,
\,+\,
c_{1\hat 3}\,\Gamma_{1}\hat\Gamma_{3}\,+\,c_{\hat 1\hat 3}\,\hat\Gamma_{13}\,
+\,c_{\hat 2\hat 3}\,\hat\Gamma_{23}\,
+\, c_{2\hat 3}\,\Gamma_{2}\hat\Gamma_{3}\,\big]\,\epsilon\,\,,
\label{DWces}
\eear
with the $c$'s  given by:
\bear
&&c_{12}\,=\,e^{2h}\,+\,\Big(\,V_{1\theta}+{a\over 2}\,\Big)
\Big(\,V_{2\phi}-{a\over 2}\,\Big)\,-\,V_{2\theta}\,V_{1\phi}\,\,,\rc\rc
&&c_{1\hat 2}\,=\,e^{h}\,\Big(\,V_{2\phi}\,-\,V_{1\theta}\,-\,a\Big)\,\,,\rc\rc
&&c_{1\hat 1}\,=\,e^{h}\,\Big(\,V_{1\phi}\,+\,V_{2\theta}\,\Big)\,\,,\rc\rc
&&c_{1\hat 3}\,=\,e^{h}\,V_{3\phi}\,\,,\rc\rc
&&c_{\hat 1\hat 3}\,=\,\Big(\,V_{1\theta}+{a\over 2}\,\Big)\,V_{3\phi}\,-\,
V_{1\phi}\,V_{3\theta}\,\,,\rc\rc
&&c_{ \hat 2\hat 3}\,=\,V_{2\theta}\,V_{3\phi}\,-\,
\Big(\,V_{2\phi}-{a\over 2}\,\Big)\,V_{3\theta}\,\,,\rc\rc
&&c_{2\hat 3}\,=\,-e^{h}\,V_{3\theta}\,\,.
\label{DWcexpression}
\eear
As mentioned at the end of section 2, we have to ensure that the kappa symmetry projection 
$\Gamma_{\kappa}\,\epsilon=\epsilon$ is compatible with the conditions (\ref{fullprojection}).
In particular,  eq. (\ref{kappacondition}) should be consistent with the second projection
written in  (\ref{fullprojection}), namely
$\Gamma_{12}\,\epsilon\,=\,\hat\Gamma_{12}\,\epsilon$. It is rather obvious that the terms in
(\ref{DWces}) containing the matrix $\hat\Gamma_3$ do not fulfil this requirement. Therefore
we must impose the vanishing of their coefficients, \ie:
\beq
c_{1\hat 3}\,=\,c_{\hat 1\hat 3}\,=\,c_{\hat 2\hat 3}\,=\,c_{ 2\hat 3}\,=\,0\,\,.
\label{DWhat3conditions}
\eeq
By inspecting the last four equations in (\ref{DWcexpression}) one readily realizes that the
conditions (\ref{DWhat3conditions}) are equivalent to:
\beq
V_{3\theta}\,=\,V_{3\phi}\,=0\,\,.
\eeq
Moreover, from the expression of $V_{3\theta}$ in (\ref{Vs}) we conclude that the condition 
$V_{3\theta}=0$ implies that
\beq
\phi_2=\phi_2(\phi_1)\,\,.
\label{phi2dep}
\eeq
Furthermore (see eq. (\ref{Vs}) ), $V_{3\phi}=0$ is equivalent to the following differential
equation:
\beq
{\partial\phi_2\over\partial\phi_1}\,=\,-{\cos\theta_1\over \cos\theta_2}\,\,.
\eeq
Let us now write
\beq
{\partial\phi_2\over\partial\phi_1}\,=\,m(\phi_1)\,\,,
\label{partialphi}
\eeq
where we have already taken into account the functional dependence written in eq.
(\ref{phi2dep}). By combining the last two equations we arrive at:
\beq
\cos\theta_2\,=\,-{\cos\theta_1\over m(\phi_1)}\,\,.
\label{costheta2}
\eeq
By differentiating eq. (\ref{costheta2})  we get
\beq
{\partial \theta_2\over \partial\theta_1}\,=\,-
{\sin\theta_1\over m(\phi_1)\sin\theta_2}\,\,.
\label{partialtheta2}
\eeq

Then, if we define
\bear
&&\Delta(\theta_1,\phi_1)\equiv{1\over 2}\,\Bigg[\,{\sin\theta_2\over \sin\theta_1}\,
\partial_{\phi_1}\phi_2\,-\,\partial_{\theta_1}\theta_2\,\Bigg]\,\,,\rc\rc
&&\tilde \Delta(\theta_1,\phi_1)\equiv{1\over 2}\,\,
{\partial_{\phi_1}\theta_2\over \sin\theta_1}\,\,,
\label{DWdeltas}
\eear
the $c$ coefficients can be written in terms of $\Delta$ and $\tilde\Delta$, namely:
\bear
&&c_{12}\,=\,e^{2h}\,-{a^2\over 4}\,-\,{1\over 4}\,+\,
{a\Delta\over 2}\,\cos\psi_0
\,-\,{a\tilde \Delta \over 2}\,\,\sin\psi_0
\,\,,\rc\rc
&&c_{1\hat 2}\,=\,e^{h}\,\big[\,\Delta\cos\psi_0
\,-\,\tilde \Delta\,\sin\psi_0\,
\,-\,a\,\big]\,\,,\rc\rc
&&c_{1\hat 1}\,=\,e^{h}\,\big[\,\Delta\,\sin\psi_0\,+\,
\tilde \Delta\,\cos\psi_0\,\,\big]\,\,,
\label{csdelta}
\eear
where we have used eqs. (\ref{phi2dep})-(\ref{partialtheta2}) and  the fact that
\beq
V_{1\theta}V_{2\phi}\,-\,V_{2\theta}V_{1\phi}\,=\,-{1\over 4}\,\,.
\eeq
Moreover,
by using the values of the derivatives $\partial_{\phi_1}\phi_2$ and 
$\partial_{\theta_1}\theta_2$ written in eqs. (\ref{partialphi}) and (\ref{partialtheta2}),
together with  eq. (\ref{costheta2}), it is easy to find 
 $\Delta(\theta_1,\phi_1)$ in terms of the function 
$m(\phi_1)$:
\beq
\Delta(\theta_1,\phi_1)\,=\,{{\rm sign}(m)\over 2}\,\,\Bigg[\,
\Bigg[1\,+\,{m(\phi_1)^2-1\over \sin^2\theta_1}\Bigg]^{{1\over 2}}
\,+\,
\Bigg[1\,+\,{m(\phi_1)^2-1\over \sin^2\theta_1}\Bigg]^{-{1\over 2}}\,\,\Bigg]\,\,,
\label{Delta}
\eeq
an expression which will be very useful in what follows.

\subsection{Abelian worldvolume solitons}
\medskip
The expression of $\Gamma_{\kappa}\,\epsilon$ that we have found above is rather complicated. In
order to tackle the general problem of finding the supersymmetric embeddings for the ansatz
(\ref{DWansatz}), let us consider the simpler problem of solving the condition
$\Gamma_{\kappa}\,\epsilon=\epsilon$ for  the abelian background, for which  $a=\alpha=0$. First
of all  let us define the following matrix:
\beq
\Gamma_{*}\equiv \Gamma_{x^0x^1x^2}\,\Gamma_r\,\Gamma_1\,\hat\Gamma_2\,\,.
\eeq
Using the fact that for the abelian background
$\Gamma_{x^0x^1x^2x^3}\,\Gamma_{12}\epsilon=\epsilon$ (see eq. (\ref{fullprojection})), one can
show that
\bear
&&\Gamma_{\kappa}\,\epsilon\,=\,{e^{3\phi}\over \sqrt{-g}}\,\sin\theta_1\,\,
\Bigg[\,\partial_rx^3\,c_{12}\,+\,c_{1\hat 2}\,\Gamma_{*}\,+\,
c_{1\hat 1}\,\hat\Gamma_{12}\,\Gamma_{*}\,+\,\rc\rc
&&\,\,\,\,\,\,\,\,\,\,\,\,\,\,\,\,\,\,\,\,\,\,\,
\,\,\,\,\,\,\,\,\,\,\,\,\,\,\,\,\,\,\,\,\,\,\,\,\,\,\,\,\,\,
\,+\,\big(\,c_{12}\,\Gamma_{*}\,+\,\partial_rx^3\,c_{1\hat 2}\,\big)\,
\Gamma_1\hat\Gamma_1\,-\,\partial_rx^3\,c_{1\hat 1}\,\Gamma_1\hat\Gamma_2\,\Bigg]\,
\epsilon\,\,.
\label{abeliankappa}
\eear
The first three terms on the right-hand side commute with the projection 
$\Gamma_{r}\hat \Gamma_{123}\,\epsilon\,=\,\epsilon$. Let us write them in detail:
\beq
\big[\partial_rx^3\,c_{12}\,+\,c_{1\hat 2}\,\Gamma_{*}\,+\,
c_{1\hat 1}\,\hat\Gamma_{12}\,\Gamma_{*}\,\big]\,\epsilon\,=
\big[\,\partial_rx^3\,c_{12}\,+\,e^h\Delta e^{\psi_0\hat\Gamma_{12}}\,\Gamma_*
\,+\,e^h\tilde\Delta \hat \Gamma_{12}\,e^{\psi_0\hat\Gamma_{12}}\,\Gamma_*\,\big]\,\epsilon\,\,.
\label{DWcomm}
\eeq
The matrix inside the brackets must act diagonally on $\epsilon$. In order to fulfil this
requirement we have to impose an extra projection to the spinor $\epsilon$. Let us define the
corresponding projector as:
\beq
{\cal P}_*\,\equiv\,\beta_1\,\Gamma_*\,+\,\beta_2\,\hat\Gamma_{12}\,\Gamma_*\,\,,
\eeq
where $\beta_1$ and $\beta_2$ are constants. We will require that $\epsilon$ satisfies the
condition:
\beq
{\cal P}_*\,\epsilon\,=\,\sigma\,\epsilon\,\,,
\label{extraDW}
\eeq
where $\sigma=\pm 1$.  For consistency ${\cal P}_*^2=1$, which, as the matrices $\Gamma_*$ and
$\hat\Gamma_{12}\,\Gamma_*$ anticommute, implies that $\beta_1^2\,+\,\beta_2^2\,=\,1$.
Accordingly, let us parametrize $\beta_1$ and $\beta_2$ in terms of a constant angle $\beta$ as
$\beta_1=\cos\beta$ and $\beta_2=\sin\beta$. The extra projection (\ref{extraDW}) takes the form:
\beq
e^{\beta\hat\Gamma_{12}}\,\Gamma_*\,\epsilon\,=\,\sigma\epsilon\,\,.
\label{extraDWbeta}
\eeq
Making use of the condition (\ref{extraDWbeta}), we can write the right-hand side of eq. 
(\ref{DWcomm}) as:
\beq
\big[\partial_rx^3\,c_{12}\,+\,e^{h}\,e^{(\psi_0\,-\,\beta)\hat\Gamma_{12}}\,\,
(\Delta\,+\,\tilde\Delta\,\hat\Gamma_{12}\,)\,\big]\,\epsilon\,\,.
\label{DWcomm2}
\eeq
We want that the matrix inside the brackets in (\ref{DWcomm2}) acts diagonally. Accordingly, we
must require that the coefficient of $\hat\Gamma_{12}$ in (\ref{DWcomm2}) vanishes which, in
turn, leads to the relation:
\beq
\tan (\beta-\psi_0)\,=\,{\tilde\Delta\over \Delta}\,\,.
\label{tanbeta}
\eeq
In particular eq. (\ref{tanbeta}) implies that $\tilde\Delta/\Delta$ must be constant. Let us
write:
\beq
{\tilde\Delta\over \Delta}\,=\,p\,=\,{\rm constant}\,\,.
\label{Deltarel}
\eeq
Let us now consider the
last three terms in (\ref{abeliankappa}), which contain matrices that do not commute with the
projection  $\Gamma_{r}\hat \Gamma_{123}\,\epsilon\,=\,\epsilon$. By using the projection 
(\ref{extraDWbeta})
these terms can be written as:
\bear
&&\bigg[\,\big(\,c_{12}\Gamma_*\,+\,\partial_rx^3\,c_{1\hat 2}\,\big)\,
\Gamma_1\hat\Gamma_1\,-\,\partial_rx^3\,c_{1\hat 1}\,\Gamma_1\hat\Gamma_2\,\,\bigg]\,
\epsilon\,=\,
\rc\rc
&&\,\,\,\,\,\,\,\,\,\,\,=\,
\bigg[\,(\partial_rx^3\,c_{1\hat 2}\,-\,\sigma c_{12}\cos\beta\,)\,
\Gamma_1\hat\Gamma_1\,+\,(\sigma c_{12}\sin\beta\,-\,\partial_rx^3\,c_{1\hat 1})\,
\Gamma_1\hat\Gamma_2\,
\bigg]\,\epsilon\,\,.
\label{noncommuting}
\eear
This contribution should vanish.  By inspecting the right-hand side of eq. 
(\ref{noncommuting}) one immediately concludes that this vanishing condition determines the
value of $\partial_rx^3$, namely:
\beq
\partial_rx^3\,=\,\sigma\,c_{12}\,{\cos\beta\over c_{1\hat 2}}\,=\,
\sigma\,c_{12}\,{\sin\beta\over c_{1\hat 1}}\,\,.
\label{partialx3first}
\eeq
The compatibility between the two expressions of $\partial_rx^3$ in eq. (\ref{partialx3first})
requires that $\tan\beta=c_{1\hat 1}/c_{1\hat 2}$. By using the values of  $c_{1\hat 1}$ and
$c_{1\hat 2}$ written in eq. (\ref{csdelta})  it is easy to verify that this compatibility
condition is equivalent to (\ref{tanbeta}). Moreover, one can write eq. (\ref{partialx3first})
as:
\beq
\partial_rx^3\,=\,{\sigma\over \Delta}\,\,e^{-h}\,\,
\big[\,e^{2h}\,-\,{1\over 4}\,\big]\,\,
{\cos\beta\over \cos\psi_0\,-\,p\sin\psi_0}\,\,.
\label{partialx3second}
\eeq
Notice that $\Delta$ only depends on the angular variables $(\theta_1,\phi_1)$.
However, since in our ansatz $x^3=x^3(r)$,  eq. (\ref{partialx3second}) is only consistent if
$\Delta$ is independent of
$(\theta_1,\phi_1)$,
\ie\ when $\Delta$ is constant. By looking at eq. (\ref{Delta}) one readily realizes that this
can only happen if $m^2=1$, \ie:
\beq
m\,=\,\pm 1\,\,.
\eeq
In this case (see eq. (\ref{Delta})) $\Delta$ is given by
\beq
\Delta=m\,\,.
\eeq
Moreover, as $\tilde\Delta\,=\,p\Delta$ (see eq. (\ref{Deltarel})), it follows that
$\tilde\Delta$ must be constant. A glance at the definition of $\tilde\Delta$ in (\ref{DWdeltas})
reveals that $\tilde\Delta$ can only be constant if it vanishes. Thus, we must have:
\beq
\tilde\Delta\,=\,0\,\,.
\eeq 
Notice that this implies that $\theta_2$ is independent of $\phi_1$ and, therefore:
\beq
\theta_2=\theta_2(\theta_1)\,\,.
\eeq
When $\tilde\Delta=0$, eq. (\ref{tanbeta}) can be solved by putting $\beta=\psi_0+n\pi$ with 
$n\in\ZZ$. Without loss of generality we can take $n=0$ or, equivalently, $\beta=\psi_0$. Then,
it follows from (\ref{extraDWbeta}) that  we must require that $\epsilon$ be an eigenvector of
$e^{\psi_0\hat\Gamma_{12}}\,\Gamma_*$, namely
\beq
e^{\psi_0\hat\Gamma_{12}}\,\Gamma_*\,\epsilon\,=\,\sigma\epsilon\,\,.
\label{DWprojection}
\eeq
Moreover, by putting $\Delta=m$, $\beta=\psi_0$ and $p=0$, eq. (\ref{partialx3second}) becomes: 
\beq
\partial_rx^3\,=\,\sigma m e^{-h}\,\Big[\,e^{2h}-{1\over 4}\,\Big]\,\,.
\label{BPSx3}
\eeq
Let us now check that the BPS equations for the embedding that we have found (eqs.
{(\ref{partialphi}) and  ({\ref{costheta2}) with $m=\pm 1$ and eq. (\ref{BPSx3})), together with
some election for the signs
$\sigma$ and $m$, are enough to guarantee the fulfilment of the kappa symmetry condition
(\ref{kappacondition}). First of all, for a general configuration with arbitrary functions
$\theta_2=\theta_2(\theta_1)$, 
$\phi_2=\phi_2(\phi_1)$ and $x^3=x^3(r)$, the determinant of the induced metric is:
\bear
&&\sqrt{-g}\,=\,e^{3\phi}\,\big[\,1\,+\,(\partial_rx^3)^2\,\big]^{{1\over 2}}\,
[\,e^{2h}\,+\,{1\over 4}\,(\partial_{\theta_1}\theta_2)^2\,\big]^{{1\over 2}}
\,\times\rc\rc
&&\,\,\,\,\,\,\,\,\,\,\,\,\,\,\,\,\,\,\,\,\,\,\,\,
\times\,\big[\,e^{2h}\sin^2\theta_1\,+\,{\cos^2\theta_1\over 4}\,+\,
{cos\theta_1\cos\theta_2\over 2}\,\partial_{\phi_1}\phi_2\,+\,
{1\over 4}\,(\partial_{\phi_1}\phi_2)^2\,\,\big]^{{1\over 2}}\,\,.\qquad
\eear
Moreover, when $x^3$ satisfies (\ref{BPSx3}), it is straightforward to prove that:
\beq
1\,+\,(\partial_rx^3)^2_{\,\,|BPS}\,=\,e^{-2h}\,\big[\,e^{2h}\,+\,{1\over 4}\,\big]^2\,\,.
\eeq
If, in addition, the angular embedding is such that  $\cos\theta_2=-m\cos\theta_1$,
$\sin\theta_2=\sin\theta_1$, 
$\partial_{\theta_1}\theta_2=-m$ with $m=\pm 1$ (see eqs. (\ref{costheta2}) and
(\ref{partialtheta2})),  one can demonstrate that:
\beq
\sqrt{-g}_{\,\,|BPS}\,=\,e^{3\phi-h}\,\sin\theta_1\,
\big[\,e^{2h}\,+\,{1\over 4}\,\big]^2\,\,.
\eeq
Moreover, in this abelian background, one can verify that:
\beq
\big[\partial_rx^3\,c_{12}\,+\,c_{1\hat 2}\,\Gamma_{*}\,+\,
c_{1\hat 1}\,\hat\Gamma_{12}\,\Gamma_{*}\,\big]
\,\epsilon_{\,\,|BPS}\,=\,\sigma m e^{-h}\,
\big[\,e^{2h}\,+\,{1\over 4}\,\big]^2\,\epsilon\,\,.
\eeq
By using these results, we see that $\Gamma_{\kappa}\epsilon=\epsilon$ if the sign $\sigma$ is
such  that
\beq
\sigma=m\,\,.
\eeq
The corresponding configurations preserve two supersymmetries, characterized by the extra
projection
\beq
e^{\psi_0\hat\Gamma_{12}}\,\Gamma_*\,\epsilon\,=\,m\epsilon\,\,,
\label{DWlastprojection}
\eeq
while $x^3(r)$ is determined by the first-order BPS differential equation
\beq
{dx^3\over dr}\,=\,e^{-h}\,\big[\,e^{2h}\,-\,{1\over 4}\,\big]\,\,.
\label{abelianx3bps}
\eeq
\subsubsection{Integration of the first-order equations}
\medskip
When $m=\pm 1$, the equations (\ref{partialphi}) and (\ref{costheta2})  that determine the
angular part of the embedding are trivial to solve. The result is:
\bear
&&\theta_2=\pi-\theta_1\,\,,
\,\,\,\,\,\,\,\,\,\,\,\,\,\,\,\,\,\,\,\,
\phi_2=\phi_1\,\,,
\,\,\,\,\,\,\,\,\,\,\,\,\,\,\,\,\,\,\,\,
(m=+1)\,\,,\rc\rc
&&\theta_2=\theta_1\,\,,
\,\,\,\,\,\,\,\,\,\,\,\,\,\,\,\,\,\,\,\,
\phi_2=2\pi-\phi_1\,\,,
\,\,\,\,\,\,\,\,\,\,\,\,\,\,\,\,\,\,\,\,
(m=-1)\,\,.
\label{DWangular}
\eear
Moreover,  by using the value of $e^{2h}$ for the abelian metric given in eq. (\ref{abelianh}),
it is also immediate to get the form of
$x^3(r)$ by direct integration of eq. (\ref{abelianx3bps}):
\beq
x^3(r)\,=\,{2\over 3}\,\,\Big(\,r-{1\over 4}\,\Big)^{{3\over 2}}\,-\,
{1\over 2}\,\,\Big(\,r-{1\over 4}\,\Big)^{{1\over 2}}\,+\,{\rm constant}\,\,.
\label{DWx3ab}
\eeq

\subsection{Non-Abelian worldvolume solitons}
\medskip
Let us now deal with the full non-abelian background. We will require that the non-abelian
solutions coincide with the abelian one in the asymptotic UV. As displayed in eq.
(\ref{epsiloneta}), the non-abelian Killing spinor $\epsilon$ is related to the asymptotic one
$\epsilon_0=f(r)\eta$ by means of a rotation 
\beq
\epsilon\,=\,e^{{\alpha \over 2}\,\Gamma_1\hat\Gamma_1}\,\,\epsilon_0\,\,,
\label{epsilon0}
\eeq
where $\alpha$ is the angle of (\ref{alpha}) and
$\epsilon_0$ satisfies the same projections as in the abelian case, namely
\beq
\Gamma_{r}\hat \Gamma_{123}\,\epsilon_0\,=\,
\Gamma_{x^0x^1x^2x^3}\,\Gamma_{12}\,\epsilon_0=\epsilon_0\,\,.
\label{proj-epsilon0}
\eeq
By using the relation between the spinors $\epsilon$ and $\epsilon_0$,
the kappa symmetry condition $\Gamma_{\kappa}\,\epsilon\,=\,\epsilon$ can be recast as a
condition on $\epsilon_0$:
\beq
e^{-{\alpha \over 2}\,\Gamma_1\hat\Gamma_1}\,\Gamma_{\kappa}\,\epsilon\,=\,\epsilon_0\,\,,
\label{kappaepsilon0}
\eeq
where the left-hand side is given by:
\bear
&&e^{-{\alpha \over 2}\,\Gamma_1\hat\Gamma_1}\,\Gamma_{\kappa}\,\epsilon\,=\,
{e^{3\phi}\over \sqrt{-g}}\,
\Gamma_{x^0x^1x^2}\,\sin\theta_1\,
\big[\,\Gamma_r\,+\,\partial_rx^3\Gamma_{x^3}\,\big]\,\times\rc\rc
&&\,\,\,\,\,\,\,\,\,\,\,\,\,\,\,\,\,\,\,\,\,\,\,\,\,\,\,\,\,\,\,
\times\big[\,
c_{12}\,e^{-\alpha\,\Gamma_1\hat\Gamma_1}\,\Gamma_{12}\,+
\,c_{1\hat 2}\,e^{-\alpha\,\Gamma_1\hat\Gamma_1}\,\Gamma_{1}\hat\Gamma_{2}\,+\, 
c_{1\hat 1}\,\Gamma_{1}\hat\Gamma_{1}\,\big]\,\epsilon_0\,\,.
\eear
Proceeding as in the abelian case, and using the projections (\ref{proj-epsilon0}), one
arrives at:
\bear
&&e^{-{\alpha \over 2}\,\Gamma_1\hat\Gamma_1}\,\Gamma_{\kappa}\,\epsilon\,=\,
{e^{3\phi}\over \sqrt{-g}}\,\sin\theta_1\,\,\Bigg[\,c_{12}\,
e^{-\alpha\,\Gamma_1\hat\Gamma_1}\,\Gamma_*\,\Gamma_1\hat\Gamma_1\,+\,
c_{1\hat 2}\,e^{-\alpha\,\Gamma_1\hat\Gamma_1}\,\Gamma_*\,+\,
c_{1\hat 1}\,\hat \Gamma_{12}\,\Gamma_*\,+\rc\rc
&&\,\,\,\,\,\,\,\,\,\,\,\,\,\,\,\,\,\,\,\,\,\,\,\,\,\,\,\,\,\,\,\,\,\,\,
+\,\,\,\partial_r x^3 c_{12}\, e^{-\alpha\,\Gamma_1\hat\Gamma_1}\,+\,
\partial_r x^3 c_{1\hat 2}\, e^{-\alpha\,\Gamma_1\hat\Gamma_1}\,\Gamma_1\hat\Gamma_1\,-\,
\partial_r x^3 c_{1\hat 1}\,\,\Gamma_1\hat\Gamma_2\,\Bigg]\,\epsilon_0\,\,.
\label{con-nocon}
\eear
In order to verify eq. (\ref{kappaepsilon0})
we shall impose to $\epsilon_0$ the same projection as in the abelian solution, namely:
\beq
e^{\psi_0\hat\Gamma_{12}}\,\Gamma_*\,\epsilon_0\,=\,\sigma\epsilon_0\,\,.
\label{asymDWprojection}
\eeq
Moreover, by expanding the exponential $e^{-\alpha\,\Gamma_1\hat\Gamma_1}$ on the
right-hand side of eq. (\ref{con-nocon}) as 
$e^{-\alpha\,\Gamma_1\hat\Gamma_1}=\cos\alpha-\sin\alpha\Gamma_1\hat\Gamma_1$ we find two
types of terms. The terms involving a matrix that commutes with the projections 
(\ref{proj-epsilon0}) are given by:
\bear
&&\bigg[\,\partial_rx^3\,(c_{12}\,\cos\alpha\,+\,c_{1\hat 2}\,\sin\alpha\,)\,+\,
(\,c_{1\hat 2}\,\cos\alpha\,-\,c_{12}\,\sin\alpha\,)\Gamma_*\,+\,
c_{1\hat 1}\,\hat\Gamma_{12}\,\Gamma_*\,\bigg]\,\epsilon_0\,\equiv\rc\rc
&&\,\,\,\,\,\,\,\,\,\,\,\,\,\,\,\,\,\,\,\,\,\,\,\,\,\,\,\,\,\,\,\,\,\,\,
\,\,\,\,\,\,\,\,\,\,\,\,\,\,\,\,\,\,\,\,\,\,\,\,\,\,\,\,\,\,\,\,\,\,\,
\equiv\,\bigg(\,{\cal A}_I\,+\,{\cal A}_{\hat 1\hat 2}\,
\hat\Gamma_{12}\,\bigg)\,\epsilon_0\,\,,
\label{calA}
\eear
while those with a matrix which does not commute with the projections are:
\bear
&&-\Gamma_1\hat\Gamma_1\,\Bigg[\,\bigg(\,
c_{12}\,\cos\alpha\,+\,c_{1\hat 2}\,\sin\alpha\,\bigg)\,\Gamma_*\,-\,
(\,c_{1\hat 2}\,\cos\alpha\,-\,c_{12}\,\sin\alpha\,)\,\partial_r x^3\,+\,
c_{1\hat 1}\,\partial_r x^3\,\hat\Gamma_{12}\,\Bigg]\epsilon_0\,=\,\rc\rc
&&\,\,\,\,\,\,\,\,\,\,\,\,\,\,\,\,\,\,\,\,\,\,\,\,\,\,\,\,\,\,\,\,\,\,\,
\,\,\,\,\,\,\,\,\,\,\,\,\,\,\,\,\,\,\,\,\,\,\,\,\,\,\,\,\,\,\,\,\,\,\,
\equiv\,-\Gamma_1\hat\Gamma_1\,
\bigg(\,{\cal B}_I\,+\,{\cal B}_{\hat 1\hat 2}\,
\hat\Gamma_{12}\,\bigg)\,\epsilon_0\,\,.
\label{calB}
\eear
The coefficients ${\cal A}$ and ${\cal B}$ defined in eqs. (\ref{calA}) and 
(\ref{calB}) can be read from the left-hand side of these equations after substituting the
value of $\Gamma_*$ from eq. (\ref{asymDWprojection}). They are given by:
\bear
&&{\cal A}_I=\partial_rx^3\,(c_{12}\,\cos\alpha\,+\,c_{1\hat 2}\,\sin\alpha\,)\,+\,
\sigma(\,c_{1\hat 2}\,\cos\alpha\,-\,c_{12}\,\sin\alpha\,)\cos\psi_0\,+\,
\sigma c_{1\hat 1}\sin\psi_0\,\,,\rc\rc
&&{\cal A}_{\hat 1\hat 2}=\sigma c_{1\hat 1}\cos\psi_0\,-\,
\sigma(\,c_{1\hat 2}\,\cos\alpha\,-\,c_{12}\,\sin\alpha\,)\sin\psi_0\,\,,\rc\rc
&&{\cal B}_I=\sigma(c_{12}\,\cos\alpha\,+\,c_{1\hat 2}\,\sin\alpha\,)\cos\psi_0\,-\,
(\,c_{1\hat 2}\,\cos\alpha\,-\,c_{12}\,\sin\alpha\,)\,\partial_rx^3\,\,,\rc\rc
&&{\cal B}_{\hat 1\hat 2}=c_{1\hat 1}\,\partial_rx^3\,-\,\sigma
(c_{12}\,\cos\alpha\,+\,c_{1\hat 2}\,\sin\alpha\,)\sin\psi_0\,\,.
\eear
 Since we are looking for
solutions which must coincide with the abelian ones in the UV, we can restrict ourselves to
the case in which $\theta_2=\theta_2(\theta_1)$, \ie\ with $\tilde\Delta=0$. It is easy to
prove that in this case  the combinations of $c_{12}$ and $c_{1\hat 2}$ appearing
above reduce to:
\bear
&&c_{12}\,\cos\alpha\,+\,c_{1\hat 2}\,\sin\alpha\,=\,\bigg[\,
r\coth 2r\,-\,{1\over 2}\,\bigg]\,\bigg[\,\coth 2r\,-\,
{\Delta\cos\psi_0\over \sinh 2r}\,\bigg]\,\,,\rc\rc
&&c_{1\hat 2}\,\cos\alpha\,-\,c_{12}\,\sin\alpha\,=\,e^h\,\bigg[\,
\Delta\cos\psi_0\coth 2r\,-\,{1\over \sinh 2r}\,\Bigg]\,\,.
\eear
To derive this result  we have used the following useful relations:
\bear
&&e^h\sin\alpha\,+\,{a\over 2}\,\cos\alpha\,=\,{1\over \sinh 2r}\,
\bigg[\,{1\over 2}\,-\,r\coth 2r\,\bigg]\,\,,\rc\rc
&&e^h\cos\alpha\,-\,{a\over 2}\,\sin\alpha\,=\,e^h\,\coth 2r\,\,,\rc
&&\bigg(\, e^{2h}\,-\,{a^2\over 4}\,-\,{1\over 4}\,\bigg)\,\sin\alpha\,+\,
ae^h\cos\alpha\,=\,{e^h\over \sinh2r}\,\,,\rc\rc
&&\bigg(\, e^{2h}\,-\,{a^2\over 4}\,-\,{1\over 4}\,\bigg)\,\cos\alpha\,-\,
ae^h\sin\alpha\,=\,\coth 2r\,\bigg[r\coth 2r\,-\,{1\over 2}\,\bigg]\,\,,
\label{MNidentities}
\eear
which can be easily demonstrated by using eqs. (\ref{MNsol}) and (\ref{alpha}). Clearly, in order
to satisfy (\ref{kappaepsilon0}) we must require that
\beq
{\cal A}_{\hat 1\hat 2}={\cal B}_I={\cal B}_{\hat 1\hat 2}=0\,\,.
\eeq
Let us now consider the  ${\cal A}_{\hat 1\hat 2}=0$   equation first. It is easy to
conclude  that this equation reduces to:
\beq
\sin\psi_0\,\bigg[\,(1-\coth 2r)
\Delta\cos\psi_0\,+\,{1\over \sinh 2r}
\,\Bigg]\,=\,0\,\,.
\eeq
If $\sin\psi_0\not=0$ the above equation can be used to obtain an expression of $\Delta$
with a non-trivial dependence on the radial variable $r$, which is in contradiction with
eq. (\ref{Delta}). Thus we conclude that $\sin\psi_0$ must vanish, \ie\ only four values
of $\psi_0$ are possible, namely:
\beq
\psi_0\,=\,0,\pi,2\pi, 3\pi\,\,.
\label{DWpsi_0}
\eeq
Let us denote
\beq
\lambda\equiv\cos\psi_0\,=\,\pm 1\,\,.
\label{DWlambda}
\eeq
Then, the condition ${\cal B}_{\hat 1\hat 2}=0$ is automatically satisfied when 
$\sin\psi_0=0$, while ${\cal B}_I=0$ leads to the following equation for $\partial_rx^3$:
\beq
\partial_rx^3\,=\,\lambda\sigma\,\,e^{-h}\,\,\,
{\cosh 2r\,-\,\Delta\lambda\over \Delta\lambda \cosh 2r\,-\,1}
\,\,\bigg[r\coth 2r\,-\,{1\over 2}\,\bigg]\,\,.
\label{DWnonabeBPSx3}
\eeq
As in the abelian case, the consistency of the above equation with our ansatz for $x^3$
requires that $\Delta$ be constant which, in turn, only can  be achieved if $m=\pm 1$ and
$\Delta=m$. Notice that this implies that the angular equations for the embedding are
exactly those written in eq. (\ref{DWangular}) for the abelian case. Moreover, when
$\theta_2=\theta_2(\theta_1)$ and $\phi_2=\phi_2(\phi_1)$ are given as in eq. 
(\ref{DWangular}), the determinant of the induced metric is
\beq
\sqrt{-g}\,=\,e^{3\phi}\,\sin\theta_1\,{r\over \sinh 2r}\,\bigg[\,
\cosh 2r\,-\,\lambda m\,\bigg]\,\sqrt{1\,+\,(\partial_r x^3)^2}\,\,.
\label{DWnonabeliandet}
\eeq
When $x^3$ satisfies the differential equation (\ref{DWnonabeBPSx3}), one can easily demonstrate
that:
\beq
\sqrt{1\,+\,(\partial_r x^3)^2}_{\,\,|BPS}\,=\,re^{-h}\
\label{x3bpsidentity}\,\,,
\eeq
and, using this result to evaluate the right-hand side of (\ref{DWnonabeliandet}), one arrives
at: 
\beq
{e^{3\phi}\,\sin\theta_1\,{\cal A}_I}_{\,\,|BPS}\,=\,\sigma m \sqrt{-g}_{\,\,|BPS}\,\,.
\eeq
Therefore,  one must take $\sigma=m$ in order to satisfy eq. (\ref{kappaepsilon0}). 
When $\sin\psi_0=0$, the
extra projection (\ref{asymDWprojection}) on the asymptotic spinor $\epsilon_0$ is
\beq
\Gamma_*\,\epsilon_0\,=\,\lambda\,m\,\epsilon_0\,\,,
\eeq
which is equivalent to the following projection on the complete spinor $\epsilon$:
\beq
e^{\alpha\Gamma_1\hat\Gamma_1}\,\Gamma_*\epsilon\,=\,\lambda\,m\,\epsilon\,\,.
\eeq
Moreover, the differential equation which determines $x^3(r)$ is:
\beq
{dx^3\over dr}\,=\,e^{-h}\,\bigg[\,r\coth 2r \,-\,{1\over 2}\,\bigg]\,\,.
\label{DWnonabeBPS}
\eeq
It is straightforward to demonstrate that this equation coincides with the abelian one in the UV.
Actually, in figure 1 we represent the result of integrating eq. (\ref{DWnonabeBPS}) and
we compare this result with that given by the function $x^3(r)$ for the abelian background (eq.
(\ref{DWx3ab})). Moreover, if we fix
the embedding $\theta_2=\theta_2(\theta_1)$, $\phi_2=\phi_2(\phi_1)$ and 
$x^3=x^3(r)$ we have two possible projections, corresponding to the two possible values
of $\lambda$. Each of these values of $\lambda$ corresponds to  two values of the angle
$\psi_0$, which again shows that the $U(1)$ symmetry of the abelian theory is broken to
$\ZZ_2$. One can check that the embeddings characterized by eqs. (\ref{DWangular}),
(\ref{DWpsi_0}) and (\ref{DWnonabeBPS}) satisfy the equations of motion derived from the
Dirac-Born-Infeld action of the probe. Moreover, it is shown in appendix A that these embeddings
saturate an energy bound, as expected for BPS worldvolume solitons.

\begin{figure}
\centerline{\epsffile{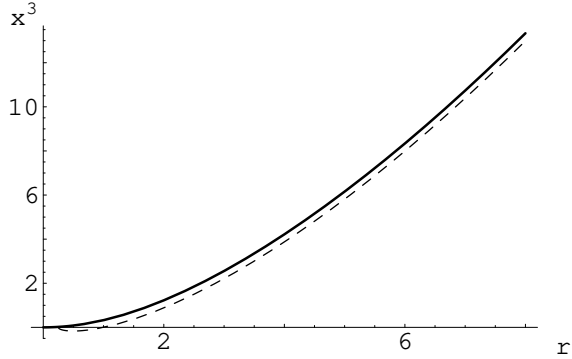}}
\caption{In this figure we represent the function $x^3(r)$ for the  wall defect in the
non-abelian background (solid line). The dashed line represents $x^3(r)$ for the abelian
background  as given by eq. (\ref{DWx3ab}). In both cases the constant of integration has been
fixed by requiring that the minimal value of $x^3$ is 0. 
}
\label{fig1}
\end{figure}

\setcounter{equation}{0}
\section{Two-dimensional defects}
\medskip
In this section we will determine BPS configurations of a D5-brane which extends along two
Minkowski coordinates (say $x^0$ and $x^1$) and along a four-dimensional submanifold embedded in
the internal part of the metric (\ref{metric}).  Such branes would be a two-dimensional
object from the gauge theory perspective and, actually, we will find that they preserve the same
supersymmetries as a D1-string stretched along $x^1$. In order to find these configurations from
the kappa symmetry condition $\Gamma_{\kappa}\,\epsilon\,=\,\epsilon$ let us choose the
following set of worldvolume coordinates for the D5-brane:
\beq
\xi^\mu\,=\,(x^0,x^1,\theta_1,\phi_1, \theta_2,\phi_2)\,\,,
\label{Stringcoordinates}
\eeq
and let us consider an embedding of the type
\beq
r\,=\,r(\theta_1,\theta_2)\,\,,
\,\,\,\,\,\,\,\,\,\,\,\,\,\,\,\,\,
\psi=\psi(\phi_1,\phi_2)\,\,,
\label{Stringansatz}
\eeq
with $x^2$ and $x^3$ being constant\footnote{For two-dimensional defects obtained with a
different election of worldvolume coordinates and ansatz, see appendix B.}. From our general
expression (\ref{GammaD5}) it is straightforward to prove that in this case
$\Gamma_{\kappa}\,\epsilon$ is given by:
\beq
\Gamma_{\kappa}\,\epsilon\,=\,{e^{\phi}\over \sqrt{-g}}\,\,\Gamma_{x^0x^1}\,
\gamma_{\theta_1\phi_1 \theta_2\phi_2}\,\epsilon\,\,.
\eeq
The induced Dirac matrices $\gamma_{\theta_i}$ and $\gamma_{\phi_i}$ are easily obtained by
using in eq. (\ref{wvgamma}) the vielbein coefficients $E_{N}^{\underline{M}}$ and our ansatz.
With the purpose of writing these matrices in a convenient form, let us define the quantities:
\beq
\Delta_i\equiv {1\over 2}\,{\cos\theta_i\,+\,\partial_{\phi_i}\psi\over \sin\theta_i}\,\,,
\label{Deltas}
\eeq
in terms of which the  $\gamma$-matrices are:
\bear
e^{-{\phi\over 2}}\,\gamma_{\theta_1}&=&e^h\Gamma_1\,+\,{a\over 2}\,\hat\Gamma_1\,+\,
\partial_{\theta_1}r\,\Gamma_r\,\,,\rc\rc
{e^{-{\phi\over 2}}\over \sin\theta_1}
\,\gamma_{\phi_1}&=&e^h\,\Gamma_2\,-\,{a\over 2}\,
\hat\Gamma_2\,+\,\Delta_1\,
\hat\Gamma_3\,\,,\rc\rc
e^{-{\phi\over 2}}\,\gamma_{\theta_2}&=&{1\over 2}\cos\psi\,\hat\Gamma_1\,-\,{1\over 2}\,
\sin\psi\,\hat\Gamma_2\,+\,\partial_{\theta_2}r\,\Gamma_r\,\,,\rc\rc
{e^{-{\phi\over 2}}\over\sin\theta_2} \,
\gamma_{\phi_2}&=&{1\over 2}\sin\psi\,\hat\Gamma_1\,+\,
{1\over 2}\cos\psi\,\hat\Gamma_2\,+\,
\Delta_2\,
\hat\Gamma_3\,\,.
\label{indgammastr}
\eear
By using eqs. (\ref{indgammastr}) and (\ref{fullprojection}) the action of the antisymmetrized
product of the $\gamma$'s on the Killing spinors $\epsilon$ can be readily obtained. It is of
the form:
\bear
{e^{-2\phi}\over \sin\theta_1\sin\theta_2}\,\,
\gamma_{\theta_1\phi_1\theta_2\phi_2}\,\epsilon\,&=&\,\big[\,b_I\,+\,
b_{2\hat 2}\,\Gamma_{2}\hat\Gamma_{2}\,+\,
b_{12}\,\Gamma_{12}\,+\,b_{1\hat 2}\,\Gamma_{1}\hat\Gamma_{2}\,+\,
b_{1\hat 3}\,\Gamma_{1}\hat\Gamma_{3}\,+\,
\rc\rc
&&+\,b_{\hat 1\hat 3}\,\hat\Gamma_{13}\,
+\,b_{\hat 2\hat 3}\,\hat\Gamma_{23}\,
+\, b_{2\hat 3}\,\Gamma_{2}\hat\Gamma_{3}\,\big]\,\epsilon\,\,,
\label{Stringces}
\eear
where the $b$'s are functions whose expression depends on the embedding of the probe. In order
to write them more compactly let us define $\Lambda_1$ and $\Lambda_2$ as follows:
\bear
&&\Lambda_1\equiv {1\over 4}\,
\Big[\,\partial_{\theta_1}r-a\cos\psi\,\partial_{\theta_2}r\,\Big]\Delta_1\,+\,
\,\Big[(\,e^{2h}-{a^2\over 4}\,)\partial_{\theta_2}r\,+\,
{a\over 4}\cos\psi\,\partial_{\theta_1}r\,\Big]\,\Delta_2\,,\rc\rc
&&\Lambda_2\equiv -{e^h\over 2}\,
\Big[\,\cos\psi\,\partial_{\theta_1}r\,-\,2a\,\partial_{\theta_2}r\,\Big]\,\Delta_2\,+\,
{e^h\over 2}\,\cos\psi\,\partial_{\theta_2}r\,\Delta_1\,\,,
\eear
where $\Delta_1$ and $\Delta_2$ have been defined in eq. (\ref{Deltas}). Then, the coefficients
of the different matrix structures appearing on the right-hand side of eq. (\ref{Stringces}) are:
\bear
&&b_I\,=\,\Lambda_1\cos\alpha\,-\,\Lambda_2\sin\alpha\,-\,
{e^{2h}\over 4}\,\,\,,\rc\rc
&&b_{2\hat 2}\,=\,\Lambda_1\sin\alpha\,+\,\Lambda_2\cos\alpha\,\,,\rc\rc
&&b_{1 2}\,=\,-\,{e^h\sin\psi\over 2}\,
\Big[\,\partial_{\theta_2}r\,\Delta_1\,-\,
\,\partial_{\theta_1}r\,\Delta_2\,\Big]\,\sin\alpha\,\,,\rc\rc
&&b_{1\hat 2}\,=\,{e^h\sin\psi\over 2}\,
\Big[\,\,\partial_{\theta_2}r\,\Delta_1\,-\,
\partial_{\theta_1}r\,\Delta_2\,\Big]\,\cos\alpha\,\,,\rc\rc
&&b_{1\hat 3}\,=\,{e^h\over 2}\,\sin\psi\Big[\,
\partial_{\theta_2}r\,(\,e^h\sin\alpha\,+\,{a\over 2}\cos\alpha\,)
\,-\,{a\over 2}\,\Delta_2\,\Big]\,\,,\rc\rc
&&b_{\hat 1\hat 3}\,=\,{e^h\over 2}\,\sin\psi\Big[\,
\,\partial_{\theta_2}r\,(\,e^h\cos\alpha\,-\,{a\over 2}\sin\alpha\,)
\,+\,e^h\,\Delta_2\,\Big]\,\,,\rc\rc
&&b_{\hat 2\hat 3}\,=\,{e^h\over 2}\,\cos\psi\Big[\,
\partial_{\theta_2}r\,(\,e^h\cos\alpha\,-\,{a\over 2}\sin\alpha\,)
\,+\,e^h\,\Delta_2\,\Big]\,+\,
{e^{h}\over 4}\,\partial_{\theta_1}r\,\sin\alpha\,\,,\rc\rc
&&b_{2\hat 3}\,=\,{e^h\over 2}\,\cos\psi
\Big[\,{a\over 2}\,\Delta_2\,-\,\partial_{\theta_2}r\,
(\,e^h\sin\alpha\,+\,{a\over 2}\cos\alpha\,)\,\Big]\,+\,
{e^h\over 4}\,\Big[\,\Delta_1\,+\,
\partial_{\theta_1}r\,\cos\alpha\,\Big]\,\,.\qquad\qquad
\eear
By inspecting the right-hand side of eq. (\ref{Stringces}) one immediately realizes that the
terms containing the matrix $\hat\Gamma_3$ give rise to contributions to $\Gamma_{\kappa}$ which
do not commute with the projection $\Gamma_{12}\,\epsilon\,=\,\hat\Gamma_{12}\,\epsilon$
satisfied by the Killing spinors (see eq. (\ref{fullprojection})). Then, if we want that the
supersymmetry preserved by the probe be compatible with that of the background, the coefficients
of these terms must vanish. Moreover, 
we would like to obtain embeddings of the D5-brane probe which preserve the same supersymmetry
as a D1-string extended along the $x^1$ direction. Accordingly\footnote{From a detailed analysis 
of the form of the $b$'s one can show that  the requirement of the vanishing of the
coefficients of the terms containing the matrix $\hat \Gamma_3$ implies the vanishing of 
$b_{2\hat 2}$, $b_{1 2}$ and $b_{1\hat 2}$. Therefore, we are not loosing generality by imposing
(\ref{cesnull}).}, we shall require the
vanishing of all terms on the right-hand side of eq. (\ref{Stringces}) except for the one
proportional to the unit matrix, \ie:
\beq
b_{2\hat 2}\,=\,b_{1 2}\,=\
b_{1\hat 2}\,=\,b_{1\hat 3}\,=\,b_{\hat 1\hat 3}\,=\,b_{\hat 2\hat 3}\,=\,
b_{2\hat 3}\,=\,0\,\,\,.
\label{cesnull}
\eeq
By plugging the explicit form of the $b$'s in (\ref{cesnull}), one gets a system of differential
equations for the embedding which will be analyzed in the rest of this section.

\subsection{Abelian worldvolume solitons}
\medskip
The above equations  (\ref{cesnull})
are quite complicated. In order to simplify the problem, let us consider first the equations
for the embedding in the abelian background, which can be obtained from the general ones by  
putting $a=\alpha=0$. In this case from  $b_{\hat 2\hat 3}=b_{2\hat 3}=0$ we get 
$\,\partial_{\theta_i} r=-\Delta_i$, where the $\Delta_i$'s have been defined in eq.
(\ref{Deltas}). More explicitly:
\beq
\partial_{\theta_i} r\,=\,-{1\over 2}\,
{\cos\theta_i+\partial_{\phi_i}\psi\over \sin\theta_i}\,\,.
\label{abelianBPS}
\eeq
One can verify that the other $b's$ in (\ref{cesnull}) vanish if these differential
equations are satisfied. Let us see the form of the kappa symmetry condition when the BPS
equations   (\ref{abelianBPS}) are satisfied. For the abelian background, the
determinant of the induced metric is given by:
\beq
\sqrt{-g}\,=\,{e^{3\phi}\over 4}\,\,\sin\theta_1\,\sin\theta_2\,
\Big[\,(\partial_{\theta_1}r)^2\,+\,4e^{2h}\,(\partial_{\theta_2}r)^2\,+\,e^{2h}\,
\Big]^{{1\over 2}}\,\,
\Big[\,\Delta_1^2\,+\,4e^{2h}\,\Delta_2^2\,+\,
e^{2h}\,\Big]^{{1\over 2}}\,\,,
\eeq
and the coefficient $b_I$ is:
\beq
b_I\,=\,{1\over 4}\,\partial_{\theta_1}r\,\Delta_1\,+\,
e^{2h}\,\partial_{\theta_2}r\,\Delta_2\,-\,
{e^{2h}\over 4}\,\,\,.
\eeq
If the BPS equations $\,\partial_{\theta_i} r=-\Delta_i$ hold, one can verify by inspection
that:
\beq
e^{-3\phi}\,\sqrt{-g}_{\,|_{BPS}}\,=\,-\sin\theta_1\sin\theta_2\,b_{I\,|_{BPS}}\,\,,
\label{sqrt-cI}
\eeq
and, thus,  the kappa symmetry condition (\ref{kappacondition}) becomes
\beq
\Gamma_{x^0x^1}\epsilon\,=-\,\epsilon\,\,,
\label{D1projection}
\eeq
which indeed corresponds to a D1 string extended along $x^1$. In this abelian case the
spinors
$\epsilon$ and $\eta$ in eq. (\ref{epsiloneta}) differ in a function which commutes with
everything. Therefore, the condition (\ref{D1projection}) translates into the same condition for
the constant spinor $\eta$, namely:
\beq
\Gamma_{x^0x^1}\eta\,=-\,\eta\,\,.
\label{D1constprojection}
\eeq
It follows that this configuration is 1/16 supersymmetric: it preserves the two supersymmetries
determined by eqs. (\ref{constantfullpro}) and (\ref{D1constprojection}).

\subsubsection{Integration of the first-order equations}
\medskip
The BPS equations (\ref{abelianBPS}) relate the partial derivatives of $r$ with those of $\psi$.
According to our ansatz (\ref{Stringansatz}) the only dependence on $\phi_1$ and $\phi_2$ in 
(\ref{abelianBPS})   comes from the derivatives of $\psi$. Therefore, 
for consistency of eq. (\ref{abelianBPS}) with our ansatz we must have:
\beq
\partial_{\phi_1}\psi\,=\,n_1\,\,,
\,\,\,\,\,\,\,\,\,\,\,\,\,\,\,\,\,\,\,\,\,\,\,\,
\partial_{\phi_2}\psi\,=\,n_2\,\,,
\eeq
where $n_1$ and $n_2$  are two constant numbers. Thus, $\psi$ must be given by:
\beq
\psi\,=\,n_1\phi_1\,+\,n_2\phi_2\,+\,{\rm constant}\,\,.
\label{winding}
\eeq
Using this form of $\psi(\phi_1,\phi_2)$ in eq. (\ref{abelianBPS}) , one can easily integrate
$r(\theta_1,\theta_2)$, namely:
\beq
e^{2r}\,=\,{C\over
\Big(\,\sin{\theta_1\over 2}\,\Big)^{n_1+1}\,\,
\Big(\,\cos{\theta_1\over 2}\,\Big)^{1-n_1}\,\,
\Big(\,\sin{\theta_2\over 2}\,\Big)^{n_2+1}\,\,
\Big(\,\cos{\theta_2\over 2}\,\Big)^{1-n_2}}\,\,,
\label{rtheta}
\eeq
where $C$ is a constant.  From the analysis of eq. (\ref{rtheta}) one easily concludes that not
all the values of the constants $n_1$ and $n_2$ are possible. Indeed, the left-hand side of eq.
(\ref{rtheta}) is always greater than one, whereas the right-hand side always vanishes for some
value of $\theta_i$ if $|n_i|>1$. Actually, we will verify in the next subsection that only when
$n_1=n_2=0$ (\ie\ when $\psi={\rm constant}$) we will be able to generalize the embedding to the
non-abelian geometry. Therefore, from now on we will concentrate only in this case, which we
rewrite as:
\beq
e^{2r}\,=\,{e^{2r_*}\over \sin\theta_1\sin\theta_2}\,\,,
\qquad\qquad
(n_1=n_2=0)\,\,,
\label{rthetazerons}
\eeq
where $r_*=r(\theta_1=\pi/2, \theta_2=\pi/2)$ is the minimal value of $r$. It is clear from
(\ref{rthetazerons}) that $r$ diverges at $\theta_i=0,\pi$. Therefore our effective strings
extend infinitely in the holographic coordinate $r$.

\subsection{Non-Abelian worldvolume solitons}
\medskip
Let us consider now the more complicated case of the non-abelian background. We are going
to argue that the kappa symmetry condition can only be solved if $\psi$ is constant and
$\sin\psi=0$. Indeed, let us assume that $\sin\psi$ does not vanish. If this is the case,
by combining the conditions $b_{\hat 1\hat 3}=0$ and $b_{\hat 2\hat 3}=0$ one gets 
$\partial_{\theta_1}r=0$. Using this result in the equation $b_{1 2}=0$, one concludes that
$\partial_{\theta_2}r=0$ (notice that the functions $\Delta_i$ can never vanish). However, 
if $r$ is independent of the $\theta_i$'s the equation $b_{ 1\hat 3}=0$ can never be
fulfilled. Thus,  we arrive at a contradiction that can only be resolved if $\sin\psi=0$.
Then, one must have:
\beq
\psi=0,\pi,2\pi,3\pi=0 \,\,(\mod\, \pi)\,\,.
\label{Strnonabpsi}
\eeq
Let us now  define
\beq
\lambda\equiv \cos\psi=\pm 1\,\,.
\label{Strlambda}
\eeq
Thus, in this non-abelian case we are only going to have zero-winding embeddings, \ie, as
anticipated above,  only the solutions with $n_1=n_2=0$ in eq. (\ref{rtheta})  generalize to the
non-abelian case. Since $\psi$ is constant, we now have
\beq
\Delta_i={1\over 2}\,\cot\theta_i\,\,.
\eeq
When $\sin\psi=0$ the equations 
$b_{1 2}=b_{1\hat 2}=b_{\hat 1\hat 3}=b_{1\hat 3}=0$ are automatically satisfied.
Moreover, the conditions $b_{\hat 2\hat 3}=b_{ 2\hat 3}=0$ reduce to:
\bear
&&\sin\alpha\,\partial_{\theta_1}r\,+\,2\lambda\,\big(\,e^h\cos\alpha\,-\,
{a\over 2}\,\sin\alpha\,\big)\,\partial_{\theta_2} r\,+\,\lambda e^h\,\cot\theta_2
\,=\,0\,\,,\rc\rc
&&\cos\alpha\,\partial_{\theta_1}r\,-\,2\lambda\,\big(\,e^h\sin\alpha\,+\,
{a\over 2}\,\cos\alpha\,\big)\,\partial_{\theta_2} r\,+\,
 {\lambda a\over 2}\,\cot\theta_2\,+\,{1\over 2}\,\cot\theta_1\,=\,0\,\,.\qquad
\label{partialsr}
\eear
From  eq. (\ref{partialsr}) one can obtain the values of the partial derivatives of $r$. Indeed,
let us define
\bear
\Delta_{\theta_1}&\equiv&{1\over 2}\,\cot\theta_1\coth(2r)\,+\,{\lambda\over 2}\,\,
{\cot\theta_2\over \sinh (2r)}\,\,,\rc\rc
\Delta_{\theta_2}&\equiv&{1\over 2}\,\cot\theta_2\coth(2r)\,+\,{\lambda\over 2}\,\,
{\cot\theta_1\over \sinh (2r)}\,\,.
\eear
Then, one has
\beq
\partial_{\theta_i}r\,=\,-\Delta_{\theta_i}\,\,.
\label{nonabelianBPS}
\eeq
To derive this result we have used  some of the identities written in eq. (\ref{MNidentities}).
Notice that $\Delta_{\theta_i}\to \Delta_i$ when $r\to\infty$ and the non-abelian
BPS equations (\ref{nonabelianBPS}) coincide with the abelian ones in eq.
(\ref{abelianBPS}) for $n_1=n_2=0$ in this limit. 
After some calculation one can check that $b_{2\hat 2}$ also vanishes as a consequence of
(\ref{nonabelianBPS}). Indeed, one can prove that $b_{2\hat 2}$ can be written:
\beq
b_{2\hat 2}\,=\,{\lambda e^h\over 2}\,\Big[\,\Delta_{\theta_1}\,\partial_{\theta_2}r\,-\,
\Delta_{\theta_2}\,\partial_{\theta_1}r\,\Big]\,\,,
\eeq
which clearly vanishes if eq. (\ref{nonabelianBPS}) is satisfied. 

For a general function $r(\theta_1,\theta_2)$, when the angle $\psi$ takes the values
written in eq. (\ref{Strnonabpsi}), the determinant of the induced metric takes the form:
\bear
&&\sqrt{-g}\,=\,{e^{3\phi}\over 4}\,\sin\theta_1\,\sin\theta_2\Big[
(\partial_{\theta_1}r)^2\,+\,4(e^{2h}\,+\,{a^2\over 4}\,)
(\partial_{\theta_2}r)^2\,-\,2a\lambda\partial_{\theta_1}r\partial_{\theta_2}r
\,+\,e^{2h}\,\Big]^{{1\over 2}}\times\rc\rc
&&\,\,\,\,\,\,\,\,\,\,\,\,\,\,\,\,\,\,\,\,\,\,\,\,\,\,\,\,\,\,\,\,\,\,\,\,\,\,
\times
\Big[\,\Delta_{\theta_1}^2\,+\,4(e^{2h}\,+\,{a^2\over 4}\,)\,
\Delta_{\theta_2}^2\,-\,2 a\lambda\Delta_{\theta_1}\Delta_{\theta_2}
\,+\,e^{2h}\,\Big]^{{1\over 2}}\,\,.
\label{sqrtg}
\eear
If the BPS equations (\ref{nonabelianBPS}) are satisfied, the two factors under the square
root on the right-hand side of eq. (\ref{sqrtg})  become equal. Moreover,   one can prove
that:
\beq
b_I\,=\,{1\over 4}\,\partial_{\theta_1}r\,(\Delta_{\theta_1}-\lambda a \Delta_{\theta_2})
\,+\,{1\over 4}\,\partial_{\theta_2}r\,
\Big(4\,(e^{2h}\,+\,{a^2\over 4}\,)\,\Delta_{\theta_2}
-\lambda a \Delta_{\theta_1}\Big)\,-\,{e^{2h}\over 4}\,\,.
\eeq
Using this result one can demonstrate, after some calculation, that eq. 
(\ref{sqrt-cI}) is also satisfied in this non-abelian case. As a consequence, the
kappa symmetry projection reduces to the one written in eq. (\ref{D1projection}), \ie\ to
that corresponding to a D1-brane.

\subsubsection{Integration of the first-order equations}
\medskip
In order to integrate the first order equations (\ref{nonabelianBPS})  for
$r(\theta_1,\theta_2)$, let us define the new variable $y(r)$ as:
\beq
y(r)\equiv \cosh(2r)\,\,.
\eeq
In terms of $y$, the BPS system (\ref{nonabelianBPS}) can be greatly simplified, namely:
\bear
\partial_{\theta_1}y\,+\,\cot\theta_1\,y&=&-\lambda\cot\theta_2\,\,,\rc\rc
\partial_{\theta_2}y\,+\,\cot\theta_2\,y&=&-\lambda\cot\theta_1\,\,,
\eear
which can be easily integrated by the method of variation of constants. In terms of the
original variable $r$ one has:
\beq
\cosh(2r)\,=\,{\cosh(2r_*)\,+\,\lambda\cos\theta_1\cos\theta_2\over
\sin\theta_1\sin\theta_2}\,\,,
\label{nonabeliansolutions}
\eeq
where $r_*\equiv r(\theta_1=\pi/2, \theta_2=\pi/2)$ is the minimal value of $r$. This is a
remarkably simple solution for the very complicated system of kappa symmetry equations.
Notice that there are two solutions for $r(\theta_1,\theta_2)$, which correspond to the two
possible values of $\lambda$ on the right-hand side of (\ref{nonabeliansolutions}). If
$\lambda=+1$ ($\lambda=-1$) the angle $\psi$ is fixed to $\psi=0,2\pi$ ($\psi=\pi,3\pi$).
Thus, the $U(1)$ symmetry $\psi\to\psi+\epsilon$ of the abelian case is broken to $\ZZ_2$,
reflecting the same breaking that occurs in the geometry. 
Moreover, it follows from (\ref{nonabeliansolutions}) that $r$ diverges at 
$\theta_{1,2}=0,\pi$. It is easily proved that the embedding written in eqs. (\ref{Strnonabpsi})
and (\ref{nonabeliansolutions}) satisfies the equations of motion of the probe and, actually, it
saturates a BPS energy bound (see appendix A). Moreover, in appendix B we will find new
codimension two defects in the non-abelian background for which the angle $\psi$ is not
constant.

\setcounter{equation}{0}
\section{Concluding remarks}
\medskip

In this paper we have systematically studied the possibility of adding supersymmetric
configurations of D5-brane probes in the MN background in such a way that they create a
codimension one or two defect in the gauge theory directions. The technique, thoroughly
explained in sections 2 and 3,  consists of using kappa symmetry to look for a system of
first-order equations which guarantee that the supersymmetry
preserved by the worldvolume of the probe is consistent with that of
the background. Although the general system of equations obtained from kappa symmetry is very
involved, the solutions we have found are remarkably simple. For a given election of worldvolume
coordinates and a given ansatz for the embedding, chosen for their simplicity and physical
significance,  the result is unique.

In order to extract consequences of our results in the gauge theory dual, some additional work
must be done. First of all, one can study the fluctuations of the probes around the
configurations found here and one can try to obtain the dictionary between these fluctuations
and the corresponding operators in the field theory side, along the lines of refs. \cite{WFO,
CEGK}. In the analysis of these fluctuations we will presumably find the difficulties associated
with the UV blowup of the dilaton, which could be overcome by using the methods employed in 
ref. \cite{flavoring} in the case of flavor branes (see also ref. \cite{Caceres:2005yx} for a
similar approach in the case of the glueball spectrum of the MN background). Once this
fluctuation-operator dictionary is obtained we could try to give some meaning to the functions
$x^3(r)$ and $r(\theta_1, \theta_2)$ of eqs. (\ref{DWnonabeBPS}) and (\ref{nonabeliansolutions})
respectively, which should encode some renormalization group flow of the defect theory. 

Let us also point out that one could explore with the same techniques employed here some other
supergravity backgrounds (such as the one obtained in \cite{MaldaNas}, which are dual to ${\cal
N}=1$, $d=3$ super Yang-Mills theory) and try  to find the configurations of probes which
introduce supersymmetric defects in the field theory. It is also worth  mentioning that, although
we have focussed here on the analysis of the supersymmetric objects in the MN background, we
could have stable non-supersymmetric configurations, such as the confining strings of ref.
\cite{Herzog:2001fq}, which are constructed from D3-branes wrapping a two-sphere. Another
example of an interesting non-supersymmetric configuration is the baryon vertex, which consists
of a D3-brane wrapped on a three-cycle which captures the RR flux \cite{Hartnoll:2004yr}.

Finally, it is also interesting to figure out which kind of  brane 
configurations of the type IIA theory correspond to the different objects studied throughout
this paper and, moreover, how they would uplift to M-theory,
working along the lines of \cite{Gomis:2001vk}.

The MN construction  of ${\cal N}=1$ Yang-Mills in the type IIB theory
by wrapping D5-branes in a finite holomorphic two-cycle of a
$CY_3$ corresponds in the type IIA theory  to D6 branes wrapping a SLag 
three cycle inside a $CY_3$, which, when going to
eleven dimensions, gets uplifted to a $G_2$ holonomy manifold.
We have seen that codimension one objects preserving
two supercharges can be added 
 by including D5-branes wrapping a
supersymmetric  three-cycle. In the type IIA theory, supersymmetric codimension one objects can
be introduced by adding a D4-brane wrapping a holomorphic 
two-cycle (this is how domain walls are introduced in ${\cal N}=1$
four dimensional field
theories, see, for instance \cite{Acharya:2001dz})
 or by considering a D6-brane wrapping a divisor
four-cycle of the $CY_3$.
 This first configuration uplifts to a M5 wrapping an associative
 three cycle inside the $G_2$ manifold, whereas the second yields
 a  cohomogeneity two $Spin(7)$ holonomy 
manifold. The two cohomogeneous
directions are the one corresponding to the energy scale and the distance
to the defect. The codimension two objects, introduced by
wrapping a D5-brane in a four-cycle yield D4-branes wrapping a susy
SLag three-cycle in the  type IIA theory, which therefore uplift to a configuration where
M5-branes wrap the coassociative four-cycle of the $G_2$ manifold.
Finally, notice that, if both kind of codimension one and two objects
are parallel, they can be
added simultaneously, preserving just one supercharge. 
The corresponding M-theory setup could consist of  a  stack of
M5-branes wrapping
a Cayley four-cycle inside the $Spin(7)$ manifold described above.
In this case, both the codimension one and two objects
are extended along the flat directions transverse to the $Spin(7)$
manifold.

\medskip
\section*{Acknowledgments}
\medskip 
We are grateful to D. Are\'an, R. Casero, J. D. Edelstein, C. N\'u\~nez  and P. Silva for
comments and discussions. The work of F. C. and A. V. R.  is supported in part by MCyT, FEDER and
Xunta de Galicia under grant  BFM2002-03881 and by  the EC Commission under the FP5 grant
HPRN-CT-2002-00325. The work of A.~P. was
partially supported by INTAS grant, 03-51-6346, CNRS PICS $\#$
2530, RTN contracts MRTN-CT-2004-005104 and MRTN-CT-2004-503369
and by a European Union Excellence Grant, MEXT-CT-2003-509661.

\vskip 1cm
\renewcommand{\theequation}{\rm{A}.\arabic{equation}}
\setcounter{equation}{0}
\medskip
\appendix

\setcounter{equation}{0}
\section{Energy bounds}
\medskip
The lagrangian density for a D5-brane probe in the Maldacena-Nu\~nez background is given by:
\beq
{\cal L}\,=\,-e^{-\phi}\,\sqrt{-g}\,-\,P[\,C^{(6)}\,]\,\,,
\label{lagrangian}
\eeq
where we have taken the string tension equal to one and $P[\,C^{(6)}\,]$ denotes the pullback of
the RR potential written in eqs. (\ref{C6}) and (\ref{calC}). In eq. (\ref{lagrangian}) we have
already taken into account that we are considering configurations of the probe with vanishing
worldvolume gauge field. For  static embeddings, such as the ones obtained in the main text, the
hamiltonian density ${\cal H}$ is just ${\cal H}= -{\cal L}$. In this appendix we are going to
verify that, for the systems studied in sections 3 and 4, ${\cal H}$ satisfies a lower  bound,
which is saturated just when the corresponding BPS equations are satisfied. Actually, we will
verify that, for a generic embedding, ${\cal H}$ can be written as:
\beq
{\cal H}\,=\,{\cal Z}\,+\,{\cal S}\,\,,
\label{H=Z+S}
\eeq
where ${\cal Z}$ is a total derivative and ${\cal S}$ is non-negative:
\beq
{\cal S}\,\ge\,0\,\,.
\label{calS}
\eeq
From eqs. (\ref{H=Z+S}) and (\ref{calS}) it follows immediately that ${\cal H}\ge {\cal Z}$,
which is the energy bound we are looking for. Moreover, we will check that 
${\cal S}=0$ precisely when the BPS equations obtained from kappa symmetry are satisfied, which
means that the energy bound is saturated for these configurations. 
These facts mean that the configurations we have found are not just solutions of the
equations of motion but BPS worldvolume solitons of the D5-brane probe \cite{GGT}.

\subsection{Energy bound for the  wall solutions}

Let us consider a D5-brane probe in the non-abelian Maldacena-Nu\~nez background and let us
choose the same worldvolume coordinates as in eq. (\ref{DWvwcoordinates}) and the ansatz
(\ref{DWansatz}) for the embedding.  For simplicity we will consider the angular embeddings
$\theta_2(\theta_1,\phi_1)$ and 
$\phi_2(\theta_1,\phi_1)$ written in  eq. (\ref{DWangular}) and we will consider a completely
arbitrary function $x^{3}(r)$. Using
the value of $\sqrt{-g}$ given in (\ref{DWnonabeliandet}), one gets:
\beq
{\cal H}\,=\,-
{\cal L}\,=\,e^{2\phi}\,\sin\theta_1\,\Bigg[\,
{r\over \sinh 2r}\,
\big(\,
\cosh 2r\,-\,\lambda m\,\big)\,\sqrt{1\,+\,(\partial_r x^3)^2}\,-\,
{\lambda m\over 4}\,a'\,\partial_r x^3\,\Bigg]\,\,,\qquad
\eeq
where $m=\pm 1$ is the same as in eq. (\ref{DWangular}) and $\lambda=\cos\psi_0\,=\,\pm 1$ (see
eq. (\ref{DWlambda})). In order to write ${\cal H}$ as in eq. (\ref{H=Z+S}) , let us
define the function 
\beq
\Lambda_r\equiv e^{-h}\,\bigg[\,r\coth 2r \,-\,{1\over 2}\,\bigg]\,\,.
\eeq
Notice that the BPS equation for $x^{3}(r)$ (eq. (\ref{DWnonabeBPS})) is just 
$\partial_r x^3=\Lambda_r$.  Furthermore, $a'$ can be written in terms
of $\Lambda_r$ as:
\beq
{a'\over 4}\,=\,-{e^h\Lambda_r\over \sinh2r}\,\,.
\eeq
Using this last result, we can write ${\cal H}$ as :
\beq
{\cal H}\,=\,\sin\theta_1\,{e^{2\phi}\over \sinh 2r}\,\Bigg[\,r\,
\big(\,\cosh 2r\,-\,\lambda m\,\big)\,\sqrt{1\,+\,(\partial_r x^3)^2}\,+\,
\lambda m e^{h}\,\Lambda_r\,\partial_r x^3\,\Bigg]\,\,.
\eeq
Let us now write ${\cal H}$ as in eq. (\ref{H=Z+S}), with:
\beq
{\cal Z}\,=\,\sin\theta_1\,{e^{2\phi+h}\over \sinh 2r}\,\Bigg[\,\cosh 2r\,
\Lambda_r\,\partial_r x^3\,+\,\cosh 2r\,-\,\lambda m\,\Bigg]\,\,.
\eeq
By using eq. (\ref{x3bpsidentity}), one can prove that
\beq
{\cal H}_{\,\,|BPS}\,=\,
{\cal Z}_{\,\,|BPS}\,\,.
\eeq
Moreover, ${\cal Z}$ can be written as a total derivative, \ie\ 
${\cal Z}\,=\,\partial_r{\cal Z}^r\,+\,\partial_{\theta_1}{\cal Z}^{\theta_1}$, with
\bear
&&{\cal Z}^r\,=\,\sin\theta_1\,{e^{2\phi+h}\over \sinh 2r}\,\Bigg[\,\cosh 2r\,
\Lambda_r\,x^3\,+\,{\sinh 2r\over 2}\,-\,\lambda m r\,\Bigg]\,\,,\rc\rc
&&{\cal Z}^{\theta_1}\,=\,\cos\theta_1\,e^{2\phi}\,\Bigg[\,2e^{2h}\,+\,
{1-a^4\over 8}\,e^{-2h}\,\Bigg]\,x^3\,\,.
\eear
To derive this result it is useful to remember that $e^{2\phi+h}/\sinh 2r$ is constant and
use the relation
\beq
\partial_r\,\bigg[\,\cosh 2r \Lambda_r\,\bigg]\,=\,e^{-h}\,\sinh 2r\,
\bigg[\,2e^{2h}\,+\,{1-a^4\over 8}\,e^{-2h}\,\bigg]\,\,,
\eeq
which can be proved by direct calculation. Moreover, taking into account that 
$r=e^{h}\,\sqrt{1+\Lambda_r^2}$ (see eq. (\ref{x3bpsidentity})), one can write ${\cal S}$
as:
\beq
{\cal S}\,=\,\sin\theta_1\,{e^{2\phi+h}\over \sinh 2r}\,
\big(\,\cosh 2r\,-\,\lambda m\,\big)\,\Bigg[\,
\sqrt{1+\Lambda_r^2}\,\sqrt{1\,+\,(\partial_r x^3)^2}\,-\,
(\,1\,+\,\Lambda_r\partial_r x^3\,)\,\Bigg]\,\,,
\eeq
and it is straightforward to verify that ${\cal S}\ge 0$ is equivalent to
\beq
(\,\partial_r x^3\,-\,\Lambda_r\,)^2\ge 0\,\,,
\eeq
which is obviously always satisfied for any function $x^3(r)$ and reduces to an equality when the
BPS equation (\ref{DWnonabeBPS}) holds.

\subsection{Energy bound for the effective string solutions}
We will now consider the configurations studied in section 4. Accordingly, let us choose
worldvolume coordinates as in (\ref{Stringcoordinates}) and an embedding of the type 
displayed in eq. (\ref{Stringansatz}) in the non-abelian MN background, where, for simplicity, we
will take the angle
$\psi$ to be a constant such that $\sin\psi=0$ (see eq. (\ref{Strnonabpsi})). In this case it is
also easy to prove that the hamiltonian density can be written as in eq. (\ref{H=Z+S}), where
for an arbitrary function $r(\theta_1,\theta_2)$, ${\cal Z}$ is a total derivative and 
${\cal S}\ge 0$. In order to verify these facts,
let us take ${\cal Z}$  to be:
\bear
&&{\cal Z}\,=\,{e^{2\phi}\over 4}\,\sin\theta_1\,\sin\theta_2\,\Big[\,
e^{2h}-(\Delta_{\theta_1}-\lambda a \Delta_{\theta_2})\partial_{\theta_1} r
\,-\,\Big(4\,(e^{2h}\,+\,{a^2\over 4}\,)\,\Delta_{\theta_2}
-\lambda a \Delta_{\theta_1}\Big)\partial_{\theta_2} r\,\Big]\,\,.\qquad\rc
\label{StrcalZ}
\eear

One can prove that ${\cal Z}$ is  a total derivative. Indeed, let us
introduce  the functions $z_1(r)$ and $z_2(r)$ as the solutions of the equations:
\bear
&&{dz_1\over dr}\,=\,\cos\alpha\,{e^{2\phi}\over 8}\,\,,\rc\rc
&&{dz_2\over dr}\,=\,-\,\Big[\,a\cos\alpha\,+\,2e^{h}\sin\alpha\,\Big]\,{e^{2\phi}\over 8}\,\,,
\eear
where $h$, $\phi$  and  $\alpha$, are the functions of the radial coordinate displayed in eqs. 
(\ref{MNsol}) and (\ref{alphaexplicit}). Then, one can verify that 
${\cal Z}\,=\,\partial_{\theta_1}\,{\cal Z}^{\theta_1}\,+\,
\partial_{\theta_2}\,{\cal Z}^{\theta_2}$, where 
\bear
{\cal Z}^{\theta_1}&=&-\cos\theta_1\sin\theta_2 \,z_1\,+\,
\lambda\sin\theta_1\cos\theta_2\,z_2\,\,,\rc\rc
{\cal Z}^{\theta_2}&=&-\sin\theta_1\cos\theta_2 \Big[\,
{e^{2\phi+2h}\over 4}\,-\,z_1\,\Big]\,-\,\lambda\cos\theta_1\sin\theta_2\,z_2\,\,.
\eear
In order to prove this result the following relation:
\beq
{d\over dr}\,\Big[\,e^{2\phi+2h}\,\Big]\,=\,2re^{2\phi}\,\,,
\eeq
is quite useful. 

It is straightforward to prove that for these configurations the pullback of $C^{(6)}$ vanishes.
Therefore (see eq. (\ref{lagrangian})), the hamiltonian density in this case is just 
${\cal H}=e^{-\phi}\,\sqrt{-g}$, with $\sqrt{-g}$ given in eq. (\ref{sqrtg}). 
Once  ${\cal Z}$ is known and given by the expression written in eq. (\ref{StrcalZ}), 
${\cal S}$ is defined as  ${\cal H}-{\cal Z}$. 
One can verify that ${\cal S}\ge 0$ is equivalent to the condition
\bear
&&\Big[\,\partial_{\theta_1} r+\Delta_{\theta_1}\,-\,\lambda a
(\partial_{\theta_2} r+\Delta_{\theta_2})\,\Big]^2\,+\,4e^{2h}\,
\Big[\,\partial_{\theta_2} r+\Delta_{\theta_2} \,\Big]^2\,+\,
4\Big[\,\Delta_{\theta_2}\partial_{\theta_1}r-\Delta_{\theta_1}\partial_{\theta_2}r
\,\Big]^2\,\ge 0\,\,,\rc
\eear
which is obviously satisfied and reduces to an identity when the BPS equations
(\ref{nonabelianBPS}) hold.  It is easy to compute the central charge ${\cal Z}$ for the BPS
configurations. The result is:
\beq
{\cal Z}_{\,|_{BPS}}\,=\,{e^{2\phi}\over 4}\,\sin\theta_1\,\sin\theta_2\,\Big[\,
(\Delta_{\theta_1}-\lambda a \Delta_{\theta_2})^2\,+\,4e^{2h}\,\Delta_{\theta_2}^2
\,+\,e^{2h}\,\Big]\,\,.
\eeq
It follows from the above expression that ${\cal Z}_{\,|_{BPS}}$ is always non-negative.

\vskip 1cm
\renewcommand{\theequation}{\rm{B}.\arabic{equation}}
\setcounter{equation}{0}
\section{More   defects}
\medskip

\subsection{Wall defects}
Let us  find more supersymmetric configurations of the D5-brane probe which
behave as a codimension one defect from the gauge theory point of view. In particular, we are
interested in trying to obtain embeddings for which the angle $\psi$ is not constant. To insure
this fact we will include $\psi$ in our set of worldvolume coordinates. Actually, we will choose
the $\xi$'s as:
\beq
\xi^{\mu}\,=\,(x^0,x^1,x^2,\theta_2,\phi_2,\psi)\,\,,
\eeq
and we will adopt the following ansatz for the embedding:
\bear
&&\theta_1=\theta_1(\theta_2),\,\,
\,\,\,\,\,\,\,\,\,\,\,\,\,
\phi_1=\phi_1(\phi_2),\,\,\rc
&&x^3=x^3(\psi)\,\,,
\,\,\,\,\,\,\,\,\,\,\,\,\,
r\,=\,r(\psi)\,\,.
\label{DW2ansatz}
\eear
For these configurations the kappa symmetry matrix
acts on the Killing spinors $\epsilon$ as:
\beq
\Gamma_{\kappa}\,\epsilon\,=\,{1\over \sqrt{-g}}\,
\gamma_{x^0 x^1x^2 \theta_2\phi_2\psi}\,\epsilon\,\,.
\eeq
Now the induced gamma matrices are:
\bear
&&e^{-{\phi\over 2}}\,\gamma_{x^{\mu}}\,=\,\Gamma_{x^{\mu}}\,\,,
\,\,\,\,\,\,\,\,\,\,\,\,\,\,\,\,\,\,
(\mu=0,1,2),\,\,\rc\rc
&&e^{-{\phi\over 2}}\,\gamma_{\theta_2}\,=\,e^h\,\partial_{\theta_2}\theta_1\,
\Gamma_{1}\,+\,W_{1\theta}\,\hat\Gamma_{1}\,+\,W_{2\theta}\,\hat\Gamma_{2}\,\,,\rc\rc
&&e^{-{\phi\over 2}}\,\gamma_{\phi_2}\,=\,e^h\,\sin\theta_1
\partial_{\phi_2}\phi_1\,
\Gamma_{2}\,+\,W_{1\phi}\,\hat\Gamma_{1}\,+\,W_{2\phi}\,\hat\Gamma_{2}\,
+\,W_{3\phi}\,\hat\Gamma_{3}\,\,,\rc\rc
&&e^{-{\phi\over 2}}\,\gamma_{\psi}\,=\,{1\over 2}\,\hat\Gamma_{3}\,+\,
\partial_{\psi}\,r\,\Gamma_r\,+\,\partial_{\psi}\,x^3\,\Gamma_{x^3}\,\,,
\label{exDWindgamma}
\eear
where the $W$'s are the following quantities:
\bear
&&W_{1\theta}\,=\,{1\over 2}\,[\cos\psi\,+\,a\partial_{\theta_2}\theta_1\,]\,\,,\rc\rc
&&W_{2\theta}\,=\,-{1\over 2}\,\sin\psi\,\,,\rc\rc
&&W_{1\phi}\,=\,{1\over 2}\,\sin\theta_2\sin\psi\,\,,\rc\rc
&&W_{2\phi}\,=\,{1\over 2}\,[\,\sin\theta_2\cos\psi\,-\,a\sin\theta_1
\partial_{\phi_2}\phi_1\,]\,\,,\rc\rc
&&W_{3\phi}\,=\,{1\over 2}\,[\cos\theta_2\,+\,\partial_{\phi_2}\phi_1
\cos\theta_1\,]\,\,.
\eear

\subsubsection{Embeddings at $r=0$}

Let us analyze first the possibility of taking in our previous equations $r=0$ and an arbitrary 
constant value of $x^3$. Since $e^h\to 0$, $a\to 1$ and $\phi\to\phi_0$ when $r\to 0$, one has in
this case $\gamma_{\theta_2}=e^{{\phi_0\over 2}}
[W_{1\theta}\,\hat\Gamma_{1}\,+\,W_{2\theta}\,\hat\Gamma_{2}]$, 
$\gamma_{\phi_2}=e^{{\phi_0\over 2}}[
W_{1\phi}\,\hat\Gamma_{1}\,+\,W_{2\phi}\,\hat\Gamma_{2}\,
+\,W_{3\phi}\,\hat\Gamma_{3}]$ and $\gamma_{\psi}=e^{{\phi_0\over 2}}\hat\Gamma_{3}/2$ and one
immediately gets:
\beq
\gamma_{\theta_2\phi_2\psi}\,\epsilon={e^{{3\over 2}\phi_0}\over 8}\,\Big[
\sin\theta_2+(\sin\theta_2\,\partial_{\theta_2}\theta_1-\sin\theta_1
\partial_{\phi_2}\phi_1)\cos\psi-\sin\theta_1\,
\partial_{\theta_2}\theta_1\,\partial_{\phi_2}\phi_1)\,\Big]
\hat\Gamma_{123}\,\epsilon\,\,.\qquad
\eeq
On the other hand, it is easy to compute the value of the determinant of the induced metric for
an embedding of the type (\ref{DW2ansatz}) at $r=0$ and constant  $x^3$. By using this result
one readily gets the action of $\Gamma_{\kappa}$ on $\epsilon$. Indeed, 
let us define $s(\theta_2,\phi_2,\psi)$ to be the following sign:
\beq
s(\theta_2,\phi_2,\psi)\equiv {\rm sign}\,
\Big[
\sin\theta_2+(\sin\theta_2\,\partial_{\theta_2}\theta_1-\sin\theta_1
\partial_{\phi_2}\phi_1)\cos\psi-\sin\theta_1\,
\partial_{\theta_2}\theta_1\,\partial_{\phi_2}\phi_1)\,\Big]\,\,.
\eeq
Then, one has:
\beq
\Gamma_{\kappa}\epsilon_{\,\,|_{r=0}}\,=\,s\,
\Gamma_{x^0x^1x^2}\hat \Gamma_{123}\epsilon_{\,\,|_{r=0}}\,\,.
\eeq
It follows that the condition $\Gamma_{\kappa}\epsilon=\epsilon$ is equivalent to the
projection:
\beq
\Gamma_{x^0x^1x^2}\hat \Gamma_{123}\epsilon_{\,\,|_{r=0}}\,=\,s\,\epsilon_{\,\,|_{r=0}}\,\,.
\label{r=0proj}
\eeq
Notice that  the right-hand side of (\ref{r=0proj}) only depends on the angular part of the
embedding through the sign $s$.  Let us rewrite eq. (\ref{r=0proj}) in  terms of the spinor
$\epsilon_0$ defined in  eq. (\ref{epsilon0}). First of all, let us introduce the matrix
$\hat\Gamma_{*}$ as:
\beq
\hat\Gamma_{*}\,=\,\Gamma_{x^0x^1x^2}\Gamma_1\hat \Gamma_{23}\,\,.
\eeq
Recall from (\ref{epsilon0}) that
$\epsilon\,=\,e^{{\alpha \over 2}\,\Gamma_1\hat\Gamma_1}\,\,\epsilon_0$. As 
$\alpha(r=0)=-\pi/2$, see eqs. (\ref{alpha}) and (\ref{alphaexplicit}),  the above condition
reduces to:
\beq
\hat\Gamma_{*}\,\epsilon_0\,=\,s\,\epsilon_0\,\,.
\label{originproj}
\eeq
It is easy to verify that this condition commutes with the projections satisfied by  
$\epsilon_0$, which are the same as those satisfied by the constant spinor $\eta$ (see eq.
(\ref{constantfullpro})). Moreover, it is readily checked that these configurations satisfy the
equations of motion of the probe. Notice that the angular embedding is undetermined. However, 
the above projection only makes sense if
$s(\theta_2,\phi_2,\psi)$ does not depend on the angles. Although the angular embedding is not
uniquely determined, there are some embeddings that can be discarded. For example if we take 
$\theta_1=\theta_2$, $\phi_1=\phi_2$ the corresponding three-cycle has vanishing volume and
$s$ is not well-defined. For $\theta_1=$ constant, $\phi_1=$ constant one has $s=1$. The same
value of $s$ is obtained if $\theta_1=\pi-\theta_2$, $\phi_1=\phi_2$ or when
$\theta_1=\theta_2$, $\phi_1=2\pi -\phi_2$.
Notice that this configuration consists of a D5-brane, which is finite in the 
internal directions, wrapping the finite $S^3$ inside the geometry, which 
has minimal volume at $r=0$. This object is thought to correspond to a 
domain wall of the field theory \cite{MN, Loewy}.
However,  the physics of domain walls is yet not fully understood in this 
model.

\subsubsection{General case}
Let us now come back to the general case. By using the relation between the spinors $\epsilon$
and $\epsilon_0$, the kappa symmetry equation  
$\Gamma_{\kappa}\,\epsilon\,=\,\epsilon$
can be rephrased as the following condition  on the spinor $\epsilon_0$:
\beq
e^{-{\alpha \over 2}\,\Gamma_1\hat\Gamma_1}\Gamma_{\kappa}
e^{{\alpha \over 2}\,\Gamma_1\hat\Gamma_1}\,\,\epsilon_0\,=\,\epsilon_0\,\,.
\label{DW2kappa-epsilon0}
\eeq
Let us evaluate the left-hand side of this equation by imposing the projection 
(\ref{originproj}), \ie\ the same projection as the one satisfied by the supersymmetric
configurations at $r=0$. After some calculation one gets an expression of the type:
\bear
&&e^{-{\alpha \over 2}\,\Gamma_1\hat\Gamma_1}\Gamma_{\kappa}
e^{{\alpha \over 2}\,\Gamma_1\hat\Gamma_1}\,\,\epsilon_0\,=\,
{e^{3\phi}\over \sqrt{-g}}\,\,\Big[\,
d_I\,+\,d_{1 \hat 1}\,\Gamma_1\hat\Gamma_1\,+\,
d_{\hat 1 \hat 2}\,\hat\Gamma_1\hat\Gamma_2\,+\,
d_{ 1 \hat 2}\,\Gamma_1\hat\Gamma_2\,+\rc\rc
&&\qquad\qquad \qquad\qquad+
d_{\hat 2 \hat 3}\,\hat\Gamma_2\hat\Gamma_3\,+\,
d_{ 2 \hat 3}\,\Gamma_2\hat\Gamma_3\,+\,
d_{\hat 1 \hat 3}\,\hat\Gamma_1\hat\Gamma_3\,+\,
d_{ 1 \hat 3}\,\Gamma_1\hat\Gamma_3\,\Big]\,\epsilon_0\,\,,
\label{Gamma-dIs}
\eear
where the $d$'s depend on the embedding (see below). Clearly, in order to satisfy
eq. (\ref{DW2kappa-epsilon0})
we must require the conditions:
\beq
d_{1 \hat 1}\,=\,d_{\hat 1 \hat 2}\,=\,d_{ 1 \hat 2}\,=\,
d_{\hat 2 \hat 3}\,=\,d_{ 2 \hat 3}\,=\,d_{\hat 1 \hat 3}\,=\,
d_{ 1 \hat 3}\,=\,0\,\,.
\eeq
The expressions of the $d$'s are quite involved. In order to write them in a compact form
let us define the quantities ${\cal P}_1$, ${\cal P}_2$ and ${\cal P}_3$ as:
\bear
&&{\cal P}_1\,\equiv\,W_{1\theta}\,W_{2\phi}\,-\,W_{1\phi}\,W_{2\theta}\,+\,
e^{2h}\,\sin\theta_1\,\partial_{\theta_2}\theta_1\,\partial_{\phi_2}\phi_1\,\,,\rc\rc
&&{\cal P}_2\,\equiv\,e^h\,\Big(\,W_{2\phi}\,\partial_{\theta_2}\theta_1\,-\,
W_{1\theta}\,\sin\theta_1\,\partial_{\phi_2}\phi_1\,\Big)\,\,,\rc\rc
&&{\cal P}_3\,\equiv\,e^h\,\Big(\,W_{1\phi}\,\partial_{\theta_2}\theta_1\,+\,
W_{2\theta}\,\sin\theta_1\,\partial_{\phi_2}\phi_1\,\Big)\,\,.
\label{Ps}
\eear
Then the coefficients of the terms that do not contain the matrix $\hat\Gamma_3$ are:
\bear
&&d_I\,=\,{s\over 2}\,\,\Big[\,{\cal P}_2\cos\alpha\,-\,{\cal P}_1\sin\alpha\,\Big]\,+\,
\Big[\,{\cal P}_1\cos\alpha\,+\,{\cal P}_2\sin\alpha\,\Big]\,\partial_{\psi}x^3\,+\,
s\,{\cal P}_3\,\partial_{\psi}r\,,\rc\rc
&&d_{1 \hat 1}\,=\,-{s\over 2}\,\,
\Big[\,{\cal P}_1\cos\alpha\,+\,{\cal P}_2\sin\alpha\,\Big]\,+\,
\Big[\,{\cal P}_2\cos\alpha\,-\,{\cal P}_1\sin\alpha\,\Big]\,\partial_{\psi}x^3\,,\rc\rc
&&d_{\hat 1 \hat 2}\,=\,{s\over 2}\,{\cal P}_3\,-\,s
\Big[\,{\cal P}_2\cos\alpha\,-\,{\cal P}_1\sin\alpha\,\Big]\,\partial_{\psi}r\,,\rc\rc
&&d_{ 1 \hat 2}\,=\,s\,
\Big[\,{\cal P}_1\cos\alpha\,+\,{\cal P}_2\sin\alpha\,\Big]\,\partial_{\psi}r\,-\,
{\cal P}_3\,\partial_{\psi}x^3\,\,.
\label{dIs}
\eear
From the conditions $d_{1 \hat 1}\,=\,d_{\hat 1 \hat 2}\,=\,0$ we get the BPS equations
that determine $\partial_{\psi}x^3$ and $\partial_{\psi}r$, namely:
\beq
\partial_{\psi}x^3\,=\,{s\over 2}\,{
{\cal P}_1\cos\alpha\,+\,{\cal P}_2\sin\alpha\over 
{\cal P}_2\cos\alpha\,-\,{\cal P}_1\sin\alpha}\,\,,
\qquad\qquad
\partial_{\psi}r\,=\,{1\over 2}\,{{\cal P}_3\over 
{\cal P}_2\cos\alpha\,-\,{\cal P}_1\sin\alpha}\,,
\label{BPSpsi}\,\,.
\eeq
while the equation $d_{ 1 \hat 2}=0$ is satisfied if the  differential equations  (\ref{BPSpsi}) 
hold.

The expressions of the coefficients of the terms with the matrix $\hat \Gamma_{3}$ are:
\bear
d_{\hat 2 \hat 3}&=&-W_{3\phi}\,\Big[\,W_{1\theta}\cos\alpha\,+\,
e^{h}\,\partial_{\theta_2}\theta_1\sin\alpha\,\Big]\,\partial_{\psi}x^3\,\,,\rc\rc
d_{ 2 \hat 3}&=&W_{3\phi}\,\Big[\,\Big(\,
W_{1\theta}\,\sin\alpha\,-\,e^{h}\,\partial_{\theta_2}\theta_1\cos\alpha\,\Big)
\partial_{\psi}x^3\,-\,s\,W_{2\theta}\,\partial_{\psi}r\,\Big]\,\,,\rc\rc
d_{ \hat 1 \hat 3}&=&W_{3\phi}\,\Big[\,W_{2\theta}\,\partial_{\psi}x^3\,-\,s
\Big(\,e^{h}\,\partial_{\theta_2}\theta_1\cos\alpha\,-\,W_{1\theta}\,\sin\alpha
\,\Big)\,\partial_{\psi}r\,\Big]\,\,,\rc\rc
d_{1 \hat 3}&=&sW_{3\phi}\,\Big[\,
W_{1\theta}\cos\alpha\,+\, e^{h}\,\partial_{\theta_2}\theta_1\sin\alpha\,\Big]\,
\partial_{\psi}r\,\,.
\label{3hat}
\eear

Let us impose now the vanishing of the coefficients (\ref{3hat}). Clearly, this condition
can be achieved by requiring that $r$ and $x^3$ be constant. It is easy to see from
the vanishing of the  right-hand side of eq. (\ref{BPSpsi})  that this only happens at 
$r=0$ and, therefore, the configuration reduces to the one studied above. Another
possibility is to impose  $W_{3\phi}=0$, which is equivalent to the following differential
equation:
\beq
-{\cos\theta_2\over \cos\theta_1}\,=\,\partial_{\phi_2}\phi_1\,\,.
\eeq
For consistency, both sides of the equation must be equal to a constant which we will denote
by $m$:
\beq
\partial_{\phi_2}\phi_1\,=\,m\,\,,
\qquad\qquad\qquad
\cos\theta_1\,=\,-{\cos\theta_2\over m}\,\,.
\label{separation}
\eeq
Moreover, 
by differentiating the above relation between $\theta_1$ and $\theta_2$, we immediately obtain:
\beq
\partial_{\theta_2}\theta_1\,=\,-{\sin\theta_2\over m
\sin\theta_1}\,=\,-{\rm sign}(m)\,
{\sin\theta_2\over \sqrt{\sin^2\theta_2\,+\,m^2-1}}\,\,.
\label{partialtheta}
\eeq
Moreover, by using eqs. (\ref{separation}) and (\ref{partialtheta}) one can easily find the
following expression of the ${\cal P}$'s:
\bear
{{\cal P}_1\over \sin\theta_2}&=&{1\over 4}\,
\Big[\,1\,+\, a\,\Big(\,\partial_{\theta_2}\theta_1\,+\,
{1\over \partial_{\theta_2}\theta_1}\,\Big)\,\cos\psi\,+\,a^2\,\Big]\,-\,e^{2h}\,\,,\rc\rc
e^{-h}\,{{\cal P}_2\over \sin\theta_2}&=&{1\over 2}\,
\Big(\,\partial_{\theta_2}\theta_1\,+\,
{1\over \partial_{\theta_2}\theta_1}\,\Big)\,\cos\psi\,+\,a\,\,,\rc\rc
e^{-h}\,{{\cal P}_3\over \sin\theta_2}&=&{1\over 2}\,
\Big(\,\partial_{\theta_2}\theta_1\,+\,
{1\over \partial_{\theta_2}\theta_1}\,\Big)\,\sin\psi
\label{BPSPs}\,\,.
\eear

For consistency with our ansatz, the right-hand side of the equation for 
$\partial_{\psi}r$ in (\ref{BPSpsi}) must necessarily be independent of $\theta_2$. By inspecting
the right-hand side of (\ref{BPSPs}) it is evident that this only happens if 
$\partial_{\theta_2}\theta_1$ is constant which, in view of eq. (\ref{partialtheta}) can
only occur if $m^2=1$, \ie\ when $m=\pm 1$. In this case $\partial_{\theta_2}\theta_1=-m$
and the angular embedding is:  
\bear
&&\theta_1=\pi-\theta_2\,\,,
\,\,\,\,\,\,\,\,\,\,\,\,\,\,\,\,\,\,\,\,
\phi_1=\phi_2\,\,,
\,\,\,\,\,\,\,\,\,\,\,\,\,\,\,\,\,\,\,\,
(m=+1)\,\,,\rc\rc
&&\theta_1=\theta_2\,\,,
\,\,\,\,\,\,\,\,\,\,\,\,\,\,\,\,\,\,\,\,
\phi_1=2\pi-\phi_2\,\,,
\,\,\,\,\,\,\,\,\,\,\,\,\,\,\,\,\,\,\,\,
(m=-1)\,\,.
\label{angularemb}
\eear
Notice that the functions in  (\ref{angularemb}) are just the same as those corresponding to the
embeddings with constant $\psi$ (eq. (\ref{DWangular})). Moreover, taking
$\partial_{\theta_2}\theta_1=-m$ in the expression of the ${\cal P}_i$'s in eq. (\ref{BPSPs}),
and substituting this result on the right-hand side of eq. (\ref{BPSpsi}), one finds the
following BPS differential equations for $r(\psi)$ and $x^3(\psi)$:

\bear
&&\partial_{\psi}r\,=\,
{1\over 2}\,{\sinh (2r)\sin\psi\over \cosh(2r)\cos\psi\,-\,m}\,\,,\rc\rc
&&\partial_{\psi}x^3\,=\,{sm\over 2}\,\,e^{-h}\,\,
\Big(\,r\coth(2r)\,-\,{1\over 2}\,\Big)\,\,
{\cosh(2r)\,-\,m\cos\psi
\over \cosh(2r)\cos\psi\,-\,m}\,\,.
\label{rx3psi}
\eear
Lets us now verify that the BPS equations written  above are enough to guarantee that 
(\ref{DW2kappa-epsilon0}) holds. With this purpose in mind, let us compute the only
non-vanishing term of the right-hand side of eq. (\ref{Gamma-dIs}), namely $d_I$.
By plugging the BPS equations (\ref{BPSpsi}) into the expression of $d_I$ in eq. (\ref{dIs}), one
gets:
\beq
{d_I}_{\,\,|BPS}\,=\,s\,\,{{\cal P}_1^2\,+\,{\cal P}_2^2\,+\,{\cal P}_3^2\over
{\cal P}_2\cos\alpha\,-\,{\cal P}_1\sin\alpha}\,\,.
\label{dIBPS}
\eeq
From  eq. (\ref{dIBPS}) one can check that:
\beq
\sqrt{-g}_{\,\,|BPS}\,=\,e^{3\phi}\,\Big|\,{d_I}_{\,\,|BPS}\,\Big|\,\,.
\eeq
In order to verify that the kappa symmetry condition (\ref{DW2kappa-epsilon0}) is satisfied we
must check that the sign of  ${d_I}_{\,\,|BPS}$ is positive. It can be verified that:
\beq
{\rm sign} \Big[\,{\cal P}_2\cos\alpha\,-\,
{\cal P}_1\sin\alpha \,]_{\,\,|BPS}\,=\,
-m\,{\rm sign} (\cos\psi)\,\,,
\eeq
and  therefore (see eq. (\ref{dIBPS})), the condition ${\rm sign}\,({d_I}_{\,\,|BPS})=+1$ holds
if the sign $s$ of the projection (\ref{originproj}) is such that:
\beq
s\,=\,-m\,\, {\rm sign}(\cos\psi)\,\,.
\label{signrelation}
\eeq
Then, given an angular embedding (\ie\ for a fixed value of $m$), we must restrict $\psi$ to a
range in which  the sign of $\cos\psi$
does not change and the sign $s$ of the projection must be chosen according to 
(\ref{signrelation}). Moreover,  one can show that the equations of
motion are satisfied if the first-order equations (\ref{rx3psi}) hold.

\subsubsection{Integration of the BPS equations}
After a short calculation one can demonstrate that
the equation for $r(\psi)$ in (\ref{rx3psi}) can be rewritten as:
\beq
\partial_{\psi}\Big[\,\cos\psi \sinh(2r)\,-\,2mr\,\Big]\,=\,0\,\,.
\eeq
In this form the BPS equation for $r(\psi)$ can be immediately integrated, namely:
\beq
\cos\psi\,=\,{C+2mr\over \sinh(2r)}\,\,,
\label{anyC}
\eeq
where $C$ is a constant. Moreover, once the function $r(\psi)$ is known, one can get $x^3(\psi)$
by direct integration of the right-hand side of the second equation in 
(\ref{rx3psi}).

Let us study the above solution for different signs of $\cos\psi$. Consider first the
region in which $\cos\psi\ge 0$, which corresponds to 
$\psi\in[-\pi/2, \pi/2]\,\,\mod\,\, 2\pi$. If the constant $C> 0$, let us represent it
in terms of a new constant $r_*$ as $C=\sinh(2r_*)-2mr_*$. Then, the above solution can be
written as:
\begin{figure}
\centerline{\epsffile{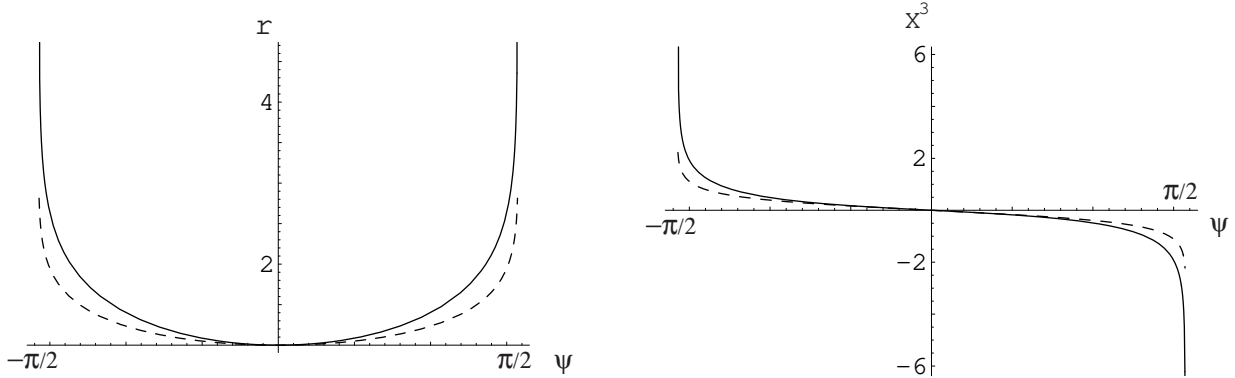}}
\caption{The functions $r(\psi)$ and $x^3(\psi)$ for the solutions (\ref{posC}) in the interval 
$\psi\in[-\pi/2, \pi/2]\,\,\mod\,\, 2\pi$. The continuous line represents the embedding with
$m=+1$, while the dashed line corresponds to $m=-1$. In this latter case $r(\psi)$ and
$x^3(\psi)$ remain finite, while for $m=+1$ they diverge at $\psi=\pm \pi/2$. }
\label{fig2}
\end{figure}
\beq
\cos\psi\,=\,{\sinh(2r_*)\,+\,2m(r-r_*)\over \sinh(2r)}\,\,,
\label{posC}
\eeq
from which it is clear that $r_*$ is the value of $r$ such that $\cos\psi=1$. The functions
$r(\psi)$ for $m=\pm 1$ written in eq. (\ref{posC}), 
and the corresponding $x^3(\psi)$, have been plotted in figure 2. 
If $m=+1$, 
the solution (\ref{posC}) is such that $r\to\infty$  and $|x^3|\to \infty$ when 
$\psi\to \pm \pi/2\,\,\mod\,\,2\pi$. However, if  $m=-1$ the radial coordinate $r$ grows
from its minimal value $r_*$ at $\psi=0\,\,\mod\,\,2\pi$ to a  maximal value 
$\hat r\,=\, r_*\,+\,{\sinh(2r_*)\over 2}$ at $\psi=\pm \pi/2\,\,\mod\,\,2\pi$, while 
$x^3(\psi)$ remains finite when $\psi\in[-\pi/2, \pi/2]\,\,\mod\,\, 2\pi$.

 If $C<0$,
it is clear from (\ref{anyC}) that, as we are considering the region  $\cos\psi\ge 0$, only
the solution with
$m=+1$ is possible. Defining $2\tilde r=-C$, the solution in this case can be written as
\beq
\cos\psi\,=\,{2(r-\tilde r)\over \sinh(2r)}\,\,,\qquad\qquad (m=+1). 
\eeq
This solution has two branches such that $r\to \tilde r, \infty$ when $\psi\to\pm\pi/2$.
Finally, if $C=0$ only the $m=+1$ solution makes sense. In this case the solution grows
from $r=0$ at $\psi=0$ to $r=\infty$ at $\psi=\pm\pi/2$.

In the region $\cos\psi\le 0$, \ie\ for  $\psi\in[\pi/2, 3\pi/2]\,\,\mod\,\, 2\pi$,
the solutions can be found from those for
$\cos\psi\ge 0$ by means of the following symmetry of the solution (\ref{anyC}):
\beq
\psi\to \pi-\psi\,\,,
\qquad\qquad
C\to-C\,\,,
\qquad\qquad
m\to-m\,\,.
\label{trasnfor}
\eeq

Then, one can get solutions in the range $\psi\in [0,2\pi]$ by joining one solution in the 
region $\cos\psi\ge 0$ to the one obtained by means of the  transformation 
(\ref{trasnfor}). Notice that the resulting solutions preserve supersymmetry at the cost
of changing the angular embedding, \ie\ by making $m\to -m$,  when the sign of $\cos\psi$
changes. In particular, in the solution obtained from the one in (\ref{posC}) when $m=-1$
the coordinate $r$ does not diverge. One can apply this construction to a single brane probe
with a singular embedding or, alternatively, one can consider two different brane probes
preserving the same supersymmetry with different angular embeddings and lying on disjoint regions
of $\psi$.

\subsection{Two-dimensional defects}

In analogy with what we have just done with the wall defect solitons, let us find some
codimension two embeddings of the D5-brane probe in which the angle $\psi$ is not constant. We
shall take the following set of worldvolume coordinates:
\beq
\xi^{\mu}\,=\,(x^0,x^1,r,\theta_2,\phi_2,\psi)\,\,,
\eeq
and we will adopt an  ansatz in which $x^2$ and $x^3$ are constant and
\beq
\theta_1=\theta_1(\theta_2),\,\,
\,\,\,\,\,\,\,\,\,\,\,\,\,
\phi_1=\phi_1(\phi_2). \,\,\rc
\eeq
The induced gamma matrices $\gamma_{x^\mu}$ $(\mu=0,1)$, $\gamma_{\theta_2}$ and 
$\gamma_{\phi_2}$ are exactly those written in eq. (\ref{exDWindgamma}), while $\gamma_{r}$ and
$\gamma_{\psi}$ are given by:
\beq
e^{-{\phi\over 2}}\,\gamma_{r}\,=\,\Gamma_r\,\,,\qquad
e^{-{\phi\over 2}}\,\gamma_{\psi}\,=\,{1\over 2}\,\hat\Gamma_{3}\,\,.
\eeq
Let us try to implement the kappa symmetry condition in the form displayed in eq.
(\ref{DW2kappa-epsilon0}). For this case, the left-hand side of (\ref{DW2kappa-epsilon0}) can be
written as:
\beq
e^{-{\alpha \over 2}\,\Gamma_1\hat\Gamma_1}\Gamma_{\kappa}
e^{{\alpha \over 2}\,\Gamma_1\hat\Gamma_1}\,\,\epsilon_0\,=\,
{e^{3\phi}\over 2\sqrt{-g}}\,\,\Gamma_{x^0 x^1}\,\Big[\,
f_I\,+\,f_{1\hat 1}\,\Gamma_1\hat\Gamma_{1}\,+\,
f_{1\hat 2}\,\Gamma_1\hat\Gamma_{2}\,\Big]\,\epsilon_0\,\,,
\eeq
where the $f$'s are expressed in terms of the ${\cal P}_i$ functions of (\ref{Ps}) as:
\bear
&&f_I\,=\,\cos\alpha\,{\cal P}_1\,+\,\sin\alpha\,{\cal P}_2\,\,,\rc\rc
&&f_{1\hat 1}\,=\,-\sin\alpha\,{\cal P}_1\,+\,\cos\alpha\,{\cal P}_2\,\,,\rc\rc
&&f_{1\hat 2}\,=\,-{\cal P}_3\,\,.
\eear
Since the matrices $\Gamma_1\hat\Gamma_1$ and $\Gamma_1\hat\Gamma_2$ do not commute with the
projection (\ref{proj-epsilon0}), it is clear that we must impose:
\beq
f_{1\hat 1}\,=\,f_{1\hat 2}\,=\,0\,\,.
\eeq
From the condition $f_{1\hat 1}\,=\,0$, we get:
\beq
{{\cal P}_2\over {\cal P}_1}\,=\,\tan\alpha\,\,,
\label{p2p1bps}
\eeq
while $f_{1\hat 2}\,=\,0$ is equivalent to the vanishing of ${\cal P}_3$, which implies:
\beq
\sin\theta_2\,\partial_{\theta_2}\theta_1\,=\,\sin\theta_1\partial_{\phi_2}\phi_1\,\,.
\label{exstringangular}
\eeq
By using this condition for the angular part of the embedding, we can write the ratio between
the functions ${\cal P}_1$  and ${\cal P}_2$ as:
\beq
{{\cal P}_2\over {\cal P}_1}\,=\,
{r\sin\alpha\,\big(\partial_{\theta_2}\theta_1\big)^2\over
r\cos\alpha\,+\,\Big(e^{2h}\,-\,{a^2\over 4}\,\Big)
\Big(\big(\partial_{\theta_2}\theta_1\big)^2-1\Big)}\,\,.
\label{p2p1bps2}
\eeq
The consistency between the expressions (\ref{p2p1bps}) and (\ref{p2p1bps2}) requires that  
$\partial_{\theta_2}\theta_1\,=\,\pm 1$. Moreover, by separating variables in the angular
embedding equation (\ref{exstringangular}) one concludes  that 
$\partial_{\phi_2}\phi_1=m$ , with  $m$ constant. Proceeding as in the previous subsection, one
easily verifies that the only consistent solutions of (\ref{exstringangular}) with 
$\partial_{\theta_2}\theta_1$ constant are:
\bear
&&\theta_1=\theta_2\,\,,
\,\,\,\,\,\,\,\,\,\,\,\,\,\,\,\,\,\,\,\,
\,\,\,\,\,\,\,\,\,\,
\phi_1=\phi_2\,\,,
\,\,\,\,\,\,\,\,\,\,\,\,\,\,\,\,\,\,\,\,
\,\,\,\,\,\,\,\,\,\,\,\,\,
(m=+1)\,\,,\rc\rc
&&\theta_1=\pi-\theta_2\,\,,
\,\,\,\,\,\,\,\,\,\,\,\,\,\,\,\,\,\,\,\,
\phi_2=2\pi-\phi_1\,\,,
\,\,\,\,\,\,\,\,\,\,\,\,\,\,\,\,\,\,\,\,
(m=-1)\,\,.
\label{Strangularemb}
\eear
Notice the difference between (\ref{Strangularemb}) and (\ref{angularemb}). One can verify that
this embedding is a solution of the equations of motion of the probe. Moreover, by computing
$\sqrt{-g}$ and $f_I$ for the embeddings (\ref{Strangularemb}), one readily proves that the
kappa symmetry condition is equivalent to the following projection on $\epsilon_0$:
\beq
\Gamma_{x^0 x^1}\,\epsilon_0\,=\,\epsilon_0\,\,.
\eeq

\end{document}